\documentclass[10pt]{article}
\usepackage{graphicx,bm,amssymb,amsmath,amsthm,xcolor}

\newcommand{\bi}{\begin{itemize}}
\newcommand{\ei}{\end{itemize}}
\newcommand{\ben}{\begin{enumerate}}
\newcommand{\een}{\end{enumerate}}
\newcommand{\be}{\begin{equation}}
\newcommand{\ee}{\end{equation}}
\newcommand{\bea}{\begin{eqnarray}} 
\newcommand{\eea}{\end{eqnarray}}
\newcommand{\ba}{\begin{align}} 
\newcommand{\ea}{\end{align}}
\newcommand{\bse}{\begin{subequations}} 
\newcommand{\ese}{\end{subequations}}
\newcommand{\bc}{\begin{center}}
\newcommand{\ec}{\end{center}}
\newcommand{\bfi}{\begin{figure}}
\newcommand{\efi}{\end{figure}}
\newcommand{\ca}[2]{\caption{#1 \label{#2}}}
\newcommand{\ig}[2]{\includegraphics[#1]{#2}}
\newcommand{\fref}[1]{{Fig.~\ref{#1}}}
\newcommand{\sref}[1]{{Sec.~\ref{#1}}}
\newcommand{\tbox}[1]{{\mbox{\tiny #1}}}
\newcommand{\mbf}[1]{{\mathbf #1}}

\newcommand{\kk}{\mbf{k}}
\newcommand{\RR}{\mathbb{R}}

\newcommand{\ttot}{{T_\tbox{tot}}}
\newcommand{\Nfea}{N_\tbox{fea}}
\newcommand{\bigO}{{\mathcal O}}
\newcommand{\ns}{\varnothing}   % no spike, empty set variant from amssymb

\DeclareMathOperator*{\argmax}{arg\,max}
\DeclareMathOperator*{\argmin}{arg\,min}
\DeclareMathOperator{\erf}{erf}
\DeclareMathOperator{\diag}{diag}
\newtheorem{thm}{Theorem}

\newtheorem{rmk}[thm]{Remark}
\begin{document}
\title{Validation of neural spike sorting algorithms
  without ground-truth information}
\author{Alex H. Barnett%
  \footnote{Simons Center for Data Analysis, and Department of Mathematics,
    Dartmouth College. {\tt ahb@math.dartmouth.edu}},
  Jeremy F. Magland%
  \footnote{Simons Center for Data Analysis, and
Department of Radiology, University of Pennsylvania},
  and
  Leslie F. Greengard%
  \footnote{Simons Center for Data Analysis, and Courant Institute, New York University}
}
\date{\today}
\maketitle
\begin{abstract}
  We describe a suite of validation metrics
  that assess the credibility of a given automatic spike sorting algorithm
  applied to a given electrophysiological recording, when ground-truth
  is unavailable.
  By rerunning the spike sorter two or more times,
  the metrics measure stability under various perturbations
  consistent with variations in the data itself,
making no assumptions about the noise model, nor
about the internal workings of the sorting algorithm.
% interface?
Such stability is a prerequisite for reproducibility of results.
We illustrate the metrics on standard sorting algorithms for
both {\em in vivo} and {\em ex vivo} recordings.
%  Assessing the accuracy of the output of a spike sorting algorithm
%  on a multielectrode neural recording currently requires much human
%  labor.
%With automated large-scale spike so
%
We believe that such metrics could
reduce the significant human labor currently spent on validation,
and should form an essential part of large-scale automated
spike sorting and systematic benchmarking of algorithms.
\end{abstract}

% IIIIIIIIIIIIIIIIIIIIIIIIIIIIIIIIIIIIIIIIIIIIIIIIIIIIIIIIIIIIIIIIIIIIIIIIII
\section{Introduction}

One of the most powerful and widely used methods for studying neuronal activity 
{\em in vivo} (for example, in behaving animals) or {\em ex vivo} (for example,
in extracted retinal preparations) is direct electrical recording. 
Using either single electrodes, tetrodes, or high-density multielectrode arrays, the 
experimental data consist of voltage patterns measured by sensors at the electrode tips, 
which are typically in the extracellular space. The strongest signals arise from an unknown 
but modest number of neurons in the immediate vicinity of each sensor,
superimposed on a background consisting substantially of signals due to more and more 
distant, electrically shielded neurons \cite{lfporigin}.
{\em Spike sorting} is the name given to algorithms that detect distinct firing events and
associate those events with specific neurons. The input consists of voltage traces from one
or more electrodes and the output is a list of 
individual neurons that have been identified, as well as the spiking (firing) times  
for each. The literature on this subject is vast,
and a very partial list of references includes
\cite{Einevoll2012,fee1996,gibson,harris2000,lewicki1998,quirogascholarpedia,quirogaopinion}.

In this environment, a typical data processing pipeline consists of (1) filtering the signal 
to remove high frequency noise and low frequency baseline drift, (2) spike detection, 
windowing and alignment, and (3) clustering based on the spike shape. 
The latter step, which is typically the core of the spike sorting algorithm, is predicated 
on the critical assumption that the electrical signals from distinct neurons have 
distinct ``shapes,"  discussed in more detail below. Finally, once the algorithm 
has assigned a neuron identity to each spike, various metrics are employed 
to assess the quality of the classification.
%\begin{rmk}
In many experiments, the signals are, in fact, {\em nonstationary}. That is, the
voltage pattern due to the firing of each neuron may vary over time. 
For the sake of simplicity, we will restrict our attention here to stationary data, but note that 
the framework for quality assessment that we introduce below does not depend on 
stationarity
%but it makes the discussion clearer.
%\end{rmk}

Unfortunately, despite major progress in algorithms and software,
spike sorting remains a labor-intensive task with
a substantial manual component, both in the clustering and quality assessment phases.
Furthermore, in the last two decades, the number of channels from which it
is becoming possible to record simultaneously has grown from one (or a few) to thousands
\cite{Einevoll2012,li2015,prentice,rossant}.
As a result, it has become an urgent matter to accelerate the data analysis, to automate
the clustering algorithms, and to develop statistical protocols that can provide robust
estimates of the accuracy of the neuronal identification.

Despite some important work on validation, discussed in the next section, however,
there are currently no established standards in the community either for estimating errors
or for assessing the reproducibility of the output of spike sorting software. 
Moreover, as noted in \cite{Hill2011,Neymotin2011,Pouzat2002,Schmitzer-Torbert2005}, 
it is of particular importance to be 
able to assess the fidelity of the output for each of the identified neurons separately.
This is vital information in subsequent modeling, since the error rate from the spike
sorting phase may permit the inclusion of some neurons and the exclusion
of others when making inferences about neural circuitry, depending on the sensitivity of the 
question being asked.

In this paper, we concentrate on the problem of quality assessment, and 
propose a framework for estimating confidence in the results obtained from a {\em black box}
spike sorting algorithm, in the absence of ground truth data such as an intra-cellular
reference electrode. 
Since rigorous estimates of accuracy are not obtainable in this
environment, we develop statistical measures of stability instead, drawing on ideas
from bootstrapping and cross-validation in statistics and machine learning
\cite{rand1971,binyu2013,zakibook,Lange2004,luxburg}. 
One of the important features of our framework is that quality estimation 
is carried out in an {\em algorithm-agnostic} fashion, not as an internal part of the spike
sorting algorithm itself.  
This requires a standard interface for the data that
is compatible with all (or most) algorithms, and the ability to invoke the spike sorting
procedure without manual intervention. Our 
validation scheme, in fact, needs to be able to re-run the spike sorting software two or more times; see \fref{f:inter}.

We hope that the stability metrics and 
neuron-by-neuron confidence measures described here will serve, in part,
as a step toward systematizing
the external evaluation of spike sorting algorithms and, in part,
as motivation to fully automate the spike sorting process. 
Indeed, we see this as crucial to progress in
the analysis of increasingly large scale electrophysiology data.
Finally, we should note that we do not claim to present a novel 
algorithm for spike sorting---we will illustrate our metrics using conventional
clustering and time series analysis tools.

The structure of this paper is as follows.
In the remainder of the introduction
we overview some existing approaches to validation (\sref{s:prior}),
then describe (\sref{s:inter})
the two versions of spike sorting that we consider:
the simpler sorting of individual spikes or ``clips''
(which have already been detected and aligned),
and the more realistic sorting of a full time series.
Standardizing the {\em interfaces} to these two algorithms is crucial
for wide applicability of our metrics.
\sref{s:clip} presents four schemes of increasing complexity for validating spike sorting on clips, and illustrates them on a standard
PCA/k-means sorting algorithm applied to {\em in vivo} rodent data.
Then \sref{s:tseries}
presents two schemes for validating the spike sorting of time series,
and illustrates them on {\em ex vivo} monkey retina data.
We discuss the implications and conclude in
\sref{s:conc}.
Some methodological and mathematical details are gathered in the two appendices.

% ppppppppppppppppppppppppppppppppppppppppppppppppppppppppppppppppppppppppp
\subsection{Prior work on quality metrics}
\label{s:prior}

As noted above, there has been substantial progress made over the last decade or so
in developing methods for assessing the output of existing spike sorting methods.
Broadly speaking, these consist of (a) statistical or information-theoretic measures of
``isolation" of the clusters of firing events associated with each neuron and
(b) physiological criteria for detecting errors such as refractory period violations.
False positives are defined as firing events that are associated with neuron $j$,
but should have been grouped with some other neuron $i$, with $i \neq j$.
False negatives are defined as firing events that should have been associated with neuron $j$,
but were either discarded or grouped with some other neuron.
We do not seek to review these approaches here, and refer the reader to 
\cite{Hill2011,Neymotin2011,Pouzat2002,Schmitzer-Torbert2005}.
%
%\co{meaning of internal/external?} % I don't get internal vs external
While the establishment of such particular metrics is
important, those of type (a) have until this point relied on
accessing {\em internal} data (eg, PCA components)
that are particular to certain types of algorithms
while excluding many other successful approaches (eg, ICA, model-based fitting, and optimal filters).
%The establishment of such internal metrics is very much complementary to the external ones developed here.
%A fair comparison of all sorting algorithms will require a metric that

Spike sorting algorithms can, of course, be run on simulated data,
where ground truth is available. Thus,
the development of more and more realistic forward
models for neuronal geometry, neuronal firing, and instrumentation noise is 
an important goal \cite{quiroga13}.
Recently another approach has emerged,
called ``surrogate'' spikes \cite{salamander}
or ``hybrid datasets'' \cite{rossant},
consisting of the original voltage recordings,
to which are added new spikes (known waveforms extracted
either from different electrodes in the same recording \cite{salamander},
or from a different recording with the same equipment \cite{rossant}),
at known times.
Running the spike sorting procedure again on the new dataset provides a powerful
validation test, particularly since it permits testing fidelity in the presence of
overlapping spikes, which tend to result in errors using simple clustering schemes.
% LG cut:
%By instead adding spikes of %precisely
%{\em all waveforms found in the
%original recording}, we extend this to become one of our
%validation metrics for time series (\sref{s:add}).

Despite all of these advances, however, most algorithms still involve manual intervention
either at late stages of the clustering process or in quality assessment.
While there have been some attempts to validate algorithms with human-operated steps
\cite{prentice}, we believe that limiting human intervention to an early, exploratory 
phase, followed by a fully automated execution of the spike sorting software will be essential
for establishing standards of reproducibility and quality, particularly with high-density
multi-electrode arrays.

The related, and broader, problem of quality metrics in clustering algorithms
has received much recent attention
\cite{Lange2004,zakibook,luxburg,clusterwise}.
% add rand1971 ?
%and we draw upon these ideas.
However, many such metrics are not directly applicable to spike sorting,
since they require access to internal variables that
may be present in some algorithms and not others,
and since they usually seek to assess the {\em overall} quality of the
clustering (eg see \cite{rand1971};
%when comparing models with different numbers of clusters.
for an exception, see \cite{clusterwise}).
In contrast, with spike sorting
the concept of ``overall quality'' is not useful:
there are often high-amplitude spikes that cluster well,
plus a ``tail'' of spikes of decreasing amplitudes that
cluster arbitrarily poorly, becoming indistinguishable from measurement
noise.
Thus, as realized by many researchers (eg \cite{Schmitzer-Torbert2005,pillow}),
per-neuron metrics (as opposed to overall performance metrics) are crucial.
% and form one of our contributions. - no, others already do it!!

\bfi[t] % Ffffffffffffffffffffffffffffffffffffffffffffffffffffffffffffffffffffff
\centering\ig{width=6.4in}{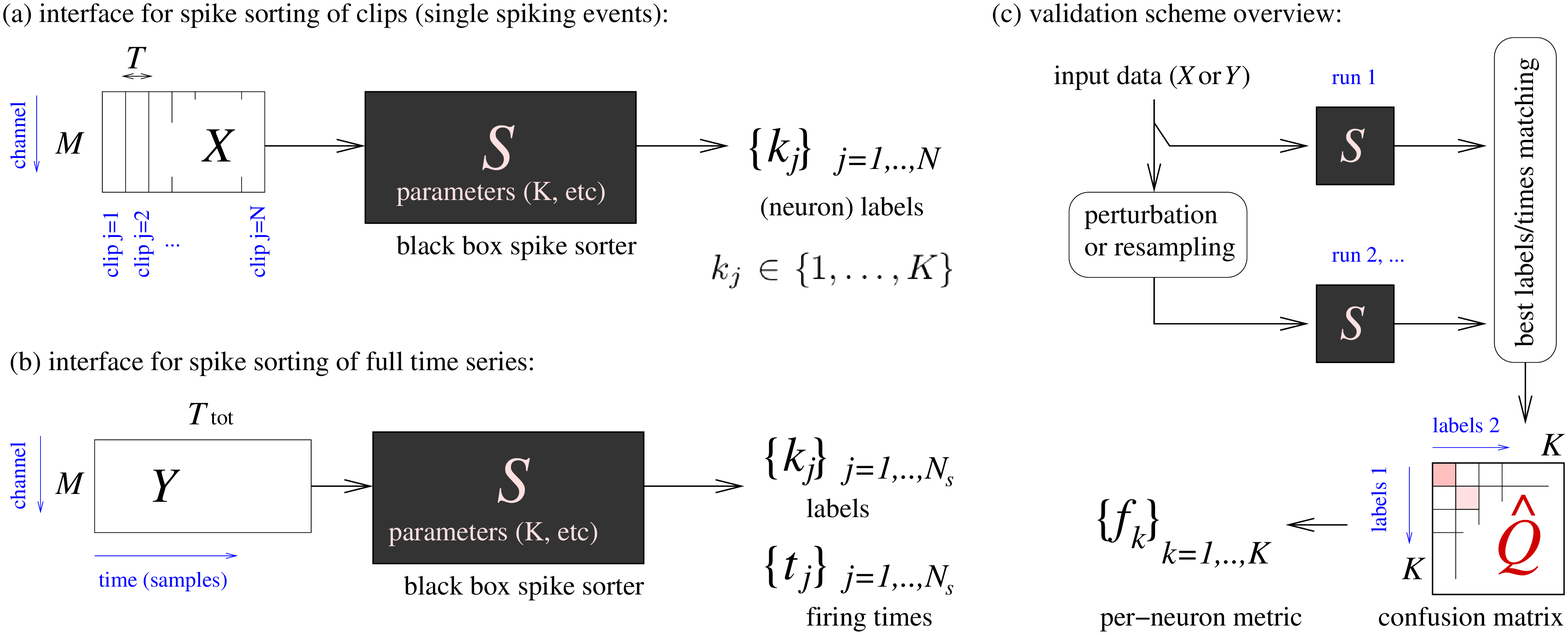}
\ca{Interfaces and validation of automatic spike sorting algorithms.
  The algorithm $S$ to be validated is inside the ``black box'', and is
  only accessible via the interface (a) or (b).
  In (a) the spike sorter assigns the $j$th clip a label (neural unit) $k_j$ from
  $1$ to $K$.
  In (b) the spike sorter acts on a whole time series $Y$ and returns
  a number $N_s$ of firing events (determined by the algorithm) described by
  labels $k_j$ and firing times $t_j$ in $[0,\ttot]$.
  (c) Overview of the validation scheme, applying to both clip-based and
  time series based algorithms, requiring the ability to rerun the
  algorithm $S$ as needed, and outputting the stability metric $f_k$.
%  The metric $f_k$ reports stability under certain variations of the data.
}{f:inter}
\efi

% iiiiiiiiiiiiiiiiiiiiiiiiiiiiiiiiiiiiiiiiiiiiiiiiiiiiiiiiiiiiiiiiiiiiiii
\subsection{Set-up of the problem and validation schemes}
\label{s:inter}

We consider {\em interfaces} to two spike sorting tasks
(\fref{f:inter}(a)--(b)).
The first is a simple classification of isolated events,
which will serve to introduce ideas,
whereas the second is a rather general spike sorting interface:
\ben
\item
The sorter classifies {\em clips}, ie short time-windows,
each presumed to contain a single spiking event.
We assume that the clips have
already been extracted (by some detection process) and aligned
(so that events of the same type occur at the same time
within the clip).
For each clip, an integer label is produced.
More formally,
let $X := \{x_{mtj}\}_{m=1,\dots,M, \; t=1,\dots,T,\; j=1,\dots,N}$
be a three-dimensional array with $N$ clips, each of duration $T$
time samples and $M$ channels (electrodes).
The sorting algorithm $S$ performs the map
\be
S(X) = \{k_j\}_{j=1}^N =: \kk
\ee
assigning a label (neuron identity) $k_j \in \{1,\dots,K\}$ to the $j$th clip.
% include $k=0$ possibility?? no
%We use the abbreviation $\kk = \{k_j\}_{j=1}^N$.
%Thus $\kk = S(X)$.
$K$ may be a fixed internal parameter of the algorithm, or may be
a variable to be determined by the algorithm.

\item
  The sorter identifies all firing times and labels
 within a time-series $Y = \{y_{mt}\}_{m=1,\dots,M,\; t=0,\dots,\ttot-1}$, ie
  \be
S(Y) = \{t_j,k_j\}_{j=1}^{N_s}
\ee
where $N_s$ is the number of spikes found (determined by the algorithm),
while $t_j\in[0,\ttot]$ and $k_j\in \{1,\dots,K\}$ are their
real-valued firing times and integer labels.
%Any detection steps (if even present) are internal to the algorithm.
%Thus the sorting algorithm is responsible for event detection
%(although it need not use such a step, see ICA Boyden REF).
In contrast to the clip-based interface,
this allows $S$ to handle overlapping spiking events,
which is crucial for complete event detection when firing rates are high
\cite{prentice,pillow,chaitu}.
\een

%We believe these interfaces encompass almost all spike-sorting
%algorithms in use.
The former interface captures the clustering task
common to much software \cite{lewicki1998,harris2000,rossant}.
The latter encompasses essentially all types of automatic sorting algorithms%
\footnote{Probabilistic (fuzzy) algorithms that output a probability density
over spike parameters such as \cite{wood,carin} could be adapted to our
interface by selecting as output
either the mode of the density, or a random sample from it.},
including those based on template matching \cite{prentice,salamander,pillow,chaitu}, ICA \cite{takahashiICA,Moore-Kochlacs2014} or
filters \cite{Franke}, that have no simple detection step.
Note that in neither interface are waveform shapes output, since
these can be reconstructed from the inputs and outputs.
%labels, or from labels and times.
We require that any algorithm parameters (thresholds, clustering settings,
electrode information, etc) are {\em fixed}, and inaccessible via the validation interface.

%stationarity. This implies that clips can re-ordered by their $j$ index
%with no meaningful effect on the outcome.
%In practice it appears that non-stationarity is uncommon for timescales
%of shorter than around 10 minutes. ref.
%This is usually enough time to collect $10^4$ or more spikes.

Our key output comparison tool is the following matrix (see \fref{f:inter}(c)).
Given two size-$N$ lists of labels, $k_j\in\{1,\dots,K\}$, $j=1,\dots,N$, and
$l_j\in\{1,\dots,L\}$, $j=1,\dots,N$,
the {\em confusion matrix} (or contingency table \cite[Ch.17]{zakibook})
is a rectangular matrix $Q$ whose $k,l$ entry is the number of
spikes labeled $k$ in the first list and $l$ in the second, ie
\be
Q_{k,l} \;:=\; \#\{j: k_j=k, l_j = l\}
~.
\label{Q}
\ee
Often the ordering of the labels is arbitrary, in which case
we seek the ``best'' permutation of one of the sets of labels,
in the sense of maximizing the sum of diagonal entries of the permuted
$Q$ matrix. This best-permuted confusion matrix $\hat Q$ is defined by
\be
\hat{Q}_{k,l} \;:=\; Q_{k,\hat\pi(l)}~,
\qquad \mbox{ where } \quad
\hat\pi \;=\; \argmax_{\pi\in S_L} \sum_{k=1}^{\min(K,L)} Q_{k,\pi(k)}
\label{Qbest}
\ee
where $S_L$ is the set of permutations of $L$ elements.
Large diagonal entries mean that the two sets of labels are consistent;
off-diagonal entries indicate the amount and type of variation between the two labelings.
The assignment problem of finding $\hat\pi$
can be solved in polynomial time by applying Kuhn's
``Hungarian algorithm'' \cite{kuhn} to $-Q$.
%
% say more about rectangular cases?
When firing times are also present, we will need an extended confusion matrix,
which we postpone until section~\ref{s:tseries}.

\bfi % ffffffffffffffffffffffffffffffffffffffffffffffffffffffffffffffffffffffffff
%(a)\ig{width=6.4in}{clips.eps}
%\\
(a)\raisebox{-1.8in}{\ig{width=1.7in}{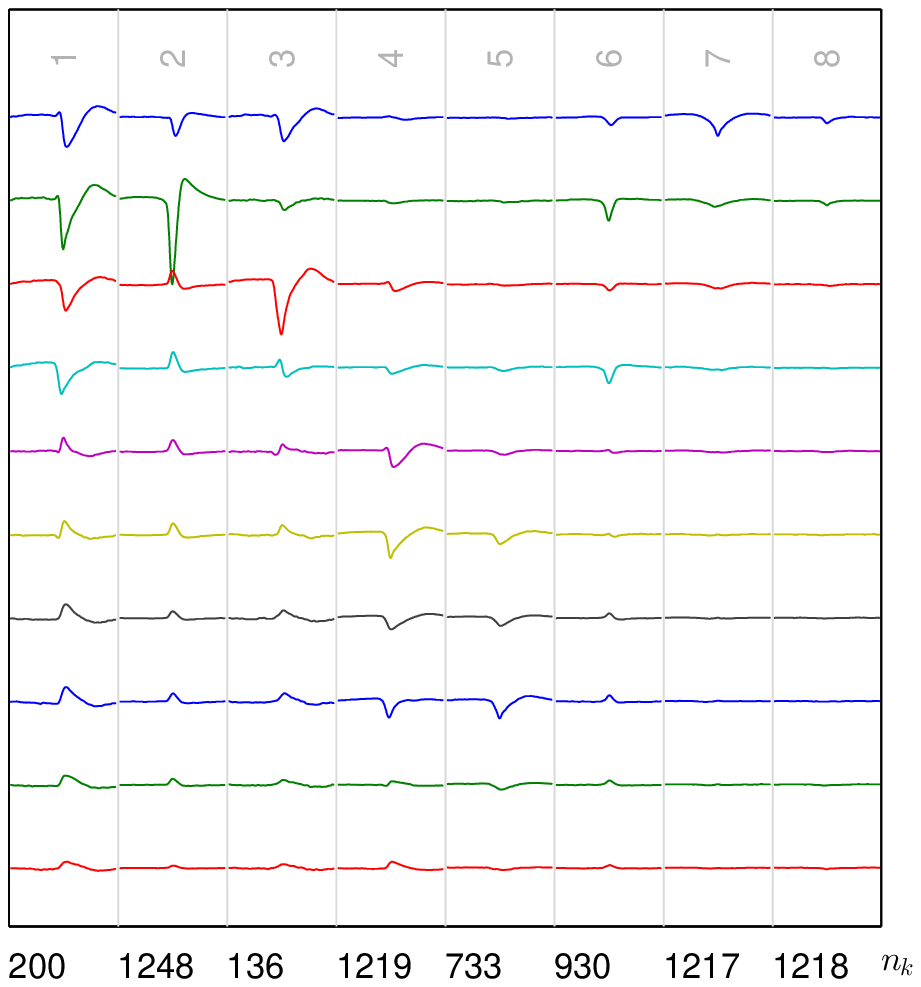}}
(b)\raisebox{-1.8in}{\ig{width=2.3in}{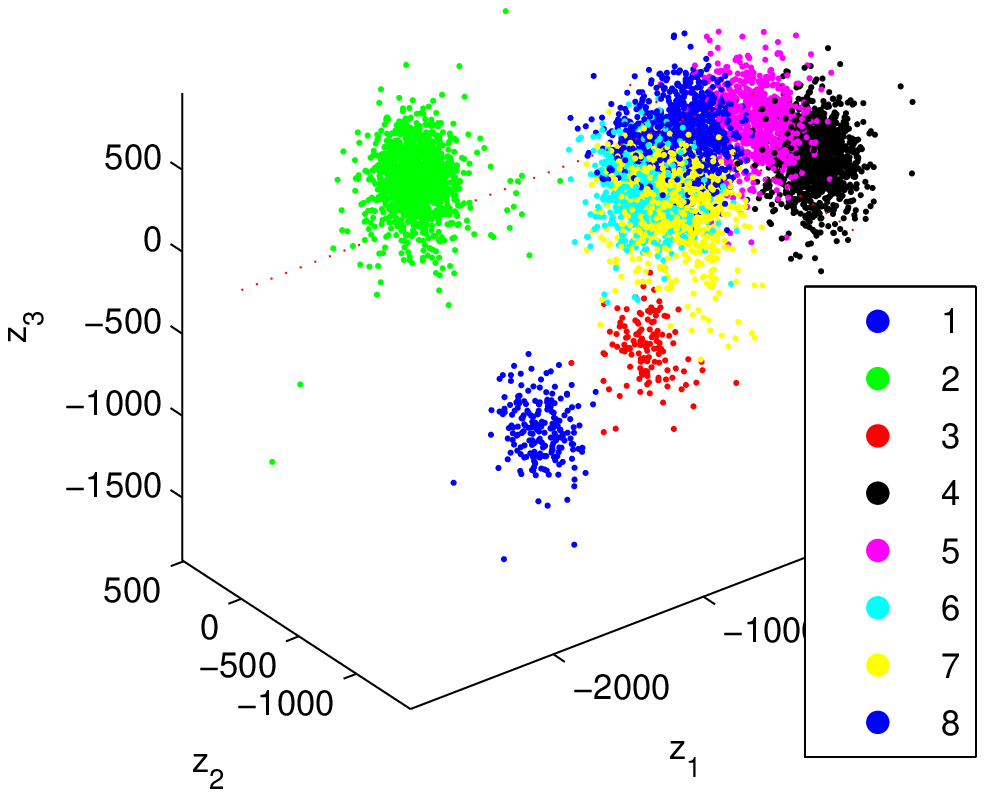}}
(c)\raisebox{-1.8in}{\ig{width=2in}{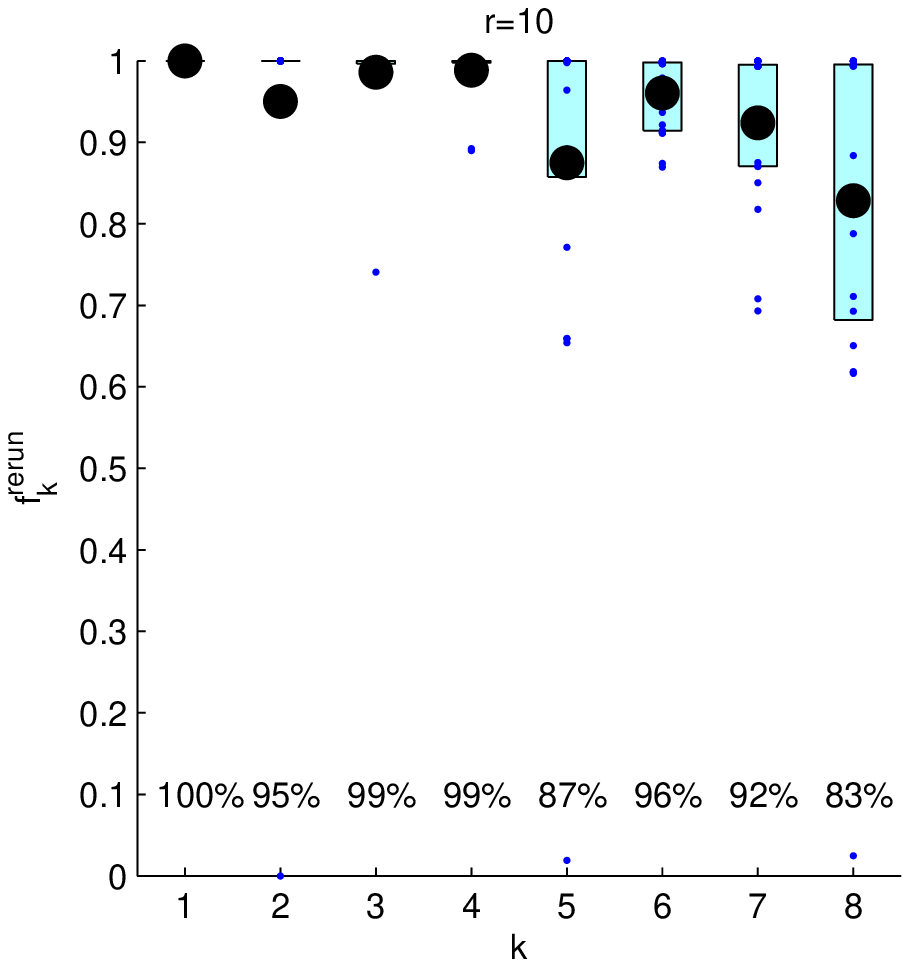}}
\ca{Spike-sorting of clips extracted from a rat motor cortex recording,
  using PCA/k-means++
  (see section~\ref{s:clip}) with $K=8$ and best of $r=100$ repeats.
%  (a) 100 example clips (chosen with roughly equal populations) grouped by their label from spike sorting via PCA/k-means++
%The duration of each clip is 2.45 ms. The label $k_j$ is shown
%below each clip.
(a) Average waveforms $W^{(k)}$ for $k=1,\dots,K$, (see \eqref{Wk}) with their
populations (numbers of firings) $n_k$ shown below.
Each of the $M=10$ channels is shown as a separate trace.
The upwards-spiking channels are a result of
spatial pre-whitening (App.~\ref{a:clips}).
%Types 1 and 4 have very small norm, and are probably associated
%with noise or multiple distant neural units.
(b) the labeling, each shown as a point in the
feature space of the first three PCA components.
Colors indicate label $k_j$ as in the inset.
%and the $z_1$ and $z_2$ axes are shown as dotted lines.
(c) 20 samples of the ``rerun'' stability metric
of Sec.~\ref{s:rerun},
for PCA/k-means++ with $K=8$ and $r=10$.
The large dot (and percentage below) shows the mean $\bar{f}^\tbox{rerun}_k$,
the small dots show all samples of $f^\tbox{rerun}_k$,
and the bar shows their 25\% to 75\% quantiles.
}{f:clips}
\efi

% CCCCCCCCCCCCCCCCCCCCCCCCCCCCCCCCCCCCCCCCCCCCCCCCCCCCCCCCCCCCCCCCCCCCCCCCCCCCC
\section{Clip-based validation schemes}
\label{s:clip}

In this section we present some clip-based validation
schemes, proceeding from simple to more sophisticated.
%
% USE THIS IN INTRO & CONCLUSIONS:
Since there is no ground truth data,
we must focus on the idea of {\em stability}.
We make the point that if the labeling of a spike is unstable
(with respect to resampling or perturbations consistent with the noise),
then it is almost certainly not accurate.
In other words, stability provides an upper bound on accuracy.
We illustrate this on a set of 6901 upsampled and peak-aligned clips extracted
from an {\em in vivo} rat motor cortex recording,
as described in Appendix~\ref{a:clips}.

\subsection{A standard clip-based spike sorting algorithm}
\label{s:ssalg}

Since our goal is to validate existing algorithms rather than
present new ones, we use as our default a standard spike-sorting algorithm
that we refer to as {\bf PCA/k-means++}, as follows.
Clips are organized into a matrix $A\in\RR^{MT\times N}$
such that each column contains the time-series for all channels for one clip.
Dimension reduction
(from dimension $MT=1470$ to dimension $\Nfea=10$ by default)
is then done by replacing each column by
its first $\Nfea$ PCA components%
\footnote{
Numerically,
$A$ is replaced by the first $\Nfea$ columns of $V^T A$,
where $V$ is the matrix whose columns are the eigenvectors of $A A^T$
ordered by descending eigenvalue.}.
Then, treating each column (clip) as a point in $\RR^{\Nfea}$, 
k-means++ \cite{Arthur07}
is used for clustering, which requires a user-specified $K$.
We remind the reader that k-means++ is a variant of k-means
using a certain favorable random initialization of the centroids.
The converged clustering produced often depends on initialization,
ie the k-means iteration is not often able to find the global minimum
of the sum of squared distances from points to their assigned centroids
(this is in general $NP$-hard \cite{Arthur07}).
Thus, as is standard, we repeat k-means++ $r$ times and choose the
repeat with the minimum sum of squared distances.
The result is a label $k_j$ for each clip.
The labels are permuted so that the $l_2$-norms of the
estimated spike waveforms, ie $(\sum_{mt} [W^{(k)}_{mt}]^2)^{1/2}$,
decrease with increasing $k$.
For this the estimated spike waveforms $W^{(k)}$ are found by simple averaging
over all clips which have been assigned the same label $k$.
More formally,
\be
W^{(k)}_{mt} = \frac{1}{n_k} \sum_{j:k_j=k} x_{mtj}~,
\qquad  m=1,\dots,M, \; t = 1,\dots,T,\; k=1,\dots,K
\label{Wk}
\ee
where
\be
n_k = n_k(\kk) := \#\{j : k_j = k\}
\label{nk}
\ee
is the $k$th population in the labeling $\kk$.

\fref{f:clips}(a)--(b) shows the result of this sorting algorithm,
for $K=8$ neurons and $r=100$ repeats, running on the clip dataset.
Note that we do not claim that this is an optimal algorithm;
%Rather, it is common algorithm, and
our goal is merely to use it to illustrate the presented validation schemes.

\bfi % ffffffffffffffffffffffffffffffffffffffffffffffffffffffffffffffffffffffffff
(a)\raisebox{-1.8in}{\ig{width=2in}{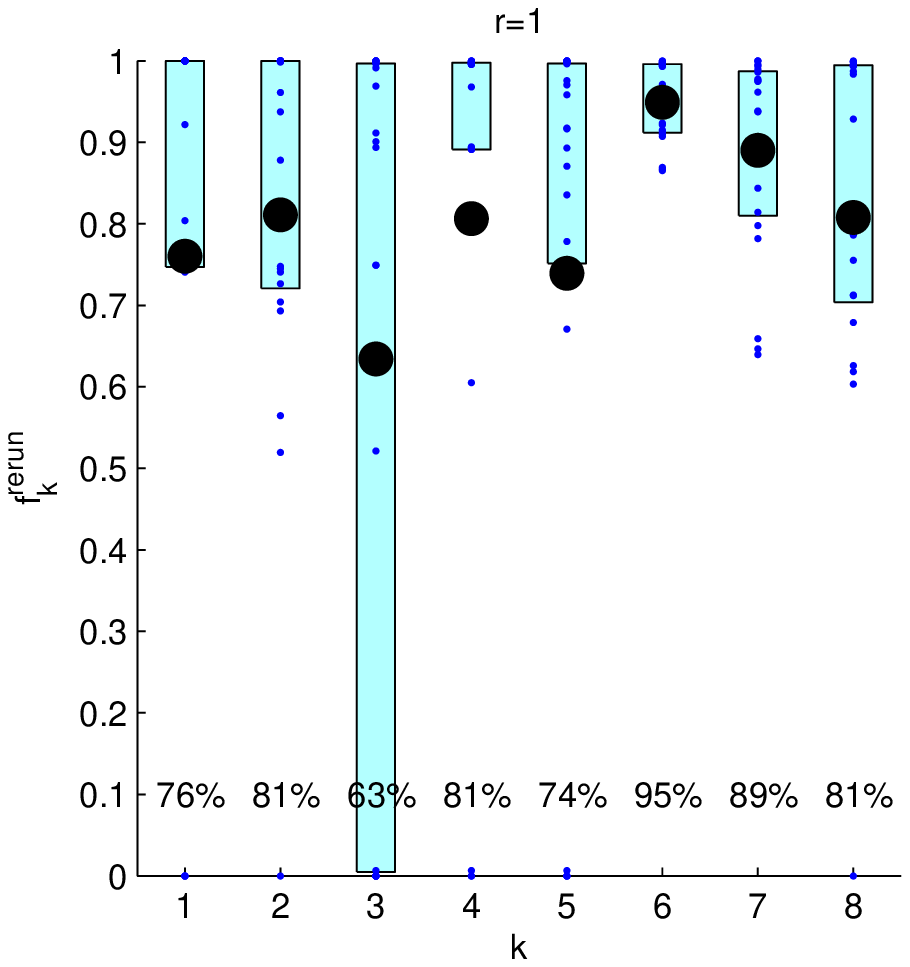}}
(b)\raisebox{-1.8in}{\ig{width=2in}{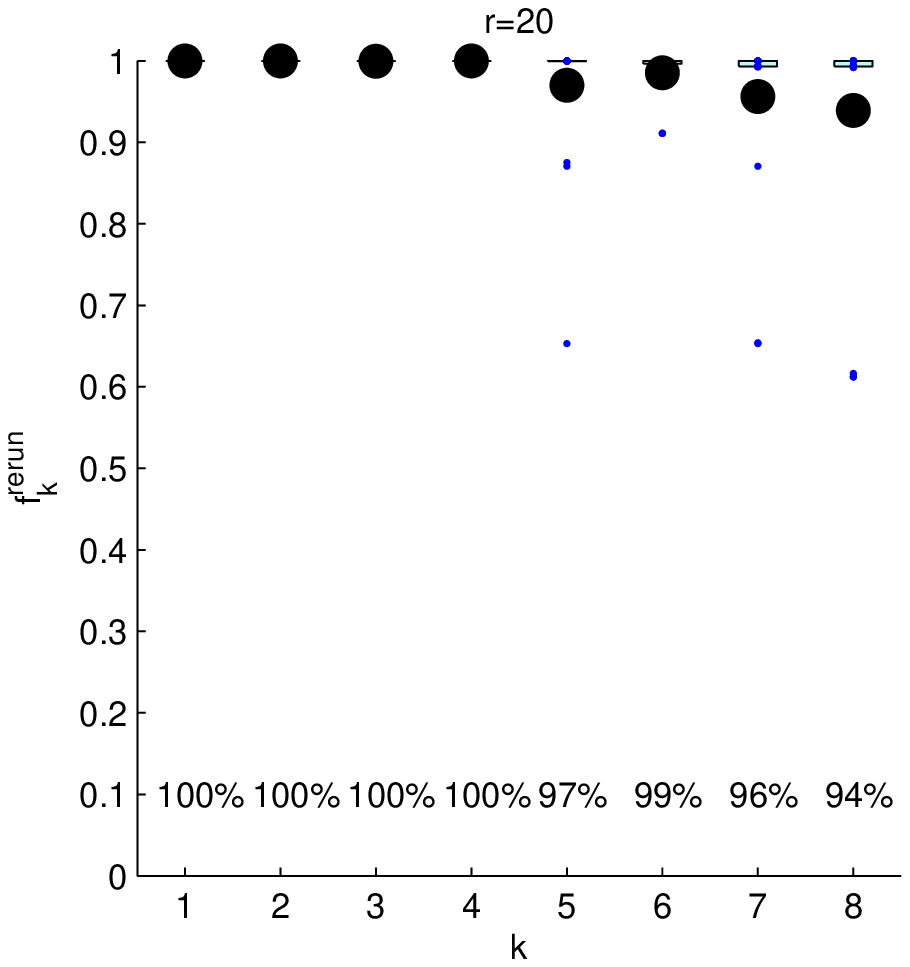}}
(c)\raisebox{-1.8in}{\ig{width=2in}{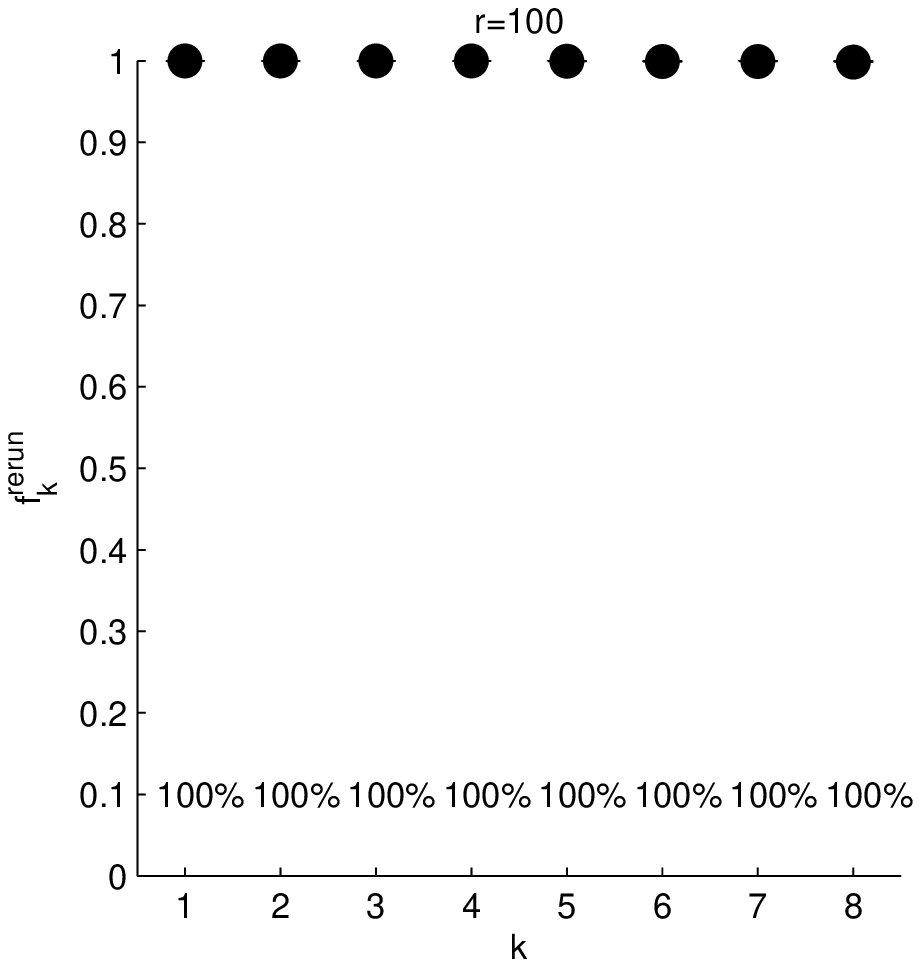}}
%\ig{width=2in}{clips_rerun_k++_Ksweep_r1.eps}
\ca{The ``rerun'' stability metric $f^\tbox{rerun}_k$
  for PCA/k-means++ with $K=8$ fixed,
  for various $r$ (best of $r$ repeats of k-means++).
  (a) $r=1$, (b) $r=20$, (c) $r=100$.
  Plots should be interpreted as in \fref{f:clips}(c).
  This shows convergence of all label stabilities to 1
  (hence an uninformative metric) as $r$ increases.
}{f:rerun_rconv}
\efi

% rrrrrrrrrrrrrrrrrrrrrrrrrrrrrrrrrrrrrrrrrrrrrrrrrrrrrrrrrrrrrrrrrrrrrrrrrrrrrrr
\subsection{A stability metric based on rerunning}
  %without subsampling or perturbation}
\label{s:rerun}

Our goal is to validate a spike sorting algorithm acting on a
particular dataset of clips.
If the spike sorter is non-deterministic,
then comparing two runs on exactly the same clip data (with different random
seeds) gives a baseline measure of reproduceability for each label.
%Indeed, it seems unwise to trust neural units from a sorter that don't enjoy
%this basic reproducibility.
Let $\{k_j\}=\kk$ and $\{l_j\}=\mbf{l}$
be the sets of labels returned from two such runs,
and assume that they have the same number of labels $K$.
Let $\hat{Q}_{k,l}$ be the elements of their best-permuted confusion
matrix \eqref{Qbest}.
Then define for each label $k$ the {\em stability},%
\footnote{This is similar to the ``F-measure'' of a clustering
  \cite[Eq.~(17.1)]{zakibook}, or harmonic mean of the precision and recall,
  relative to a ground truth clustering,
  with the difference that a
  single matching is enforced between the two sets of labels.}
\be
f_k := \frac{2\hat{Q}_{k,k}}{n_k(\kk) + n_k(\mbf{l})}
~,\qquad k=1,\dots, K.
\label{fk}
\ee
Samples of $f_k$ are now computed by performing this process
for many independent pairs of runs, with the same $K$,
and the mean $\bar{f}_k$ is reported.
%(following the clustering literature \cite{luxburg})
We use the notation $f^\tbox{rerun}_k$ to indicate the ``rerun'' stability
metric. Note that the pairs of runs used to sample the metric
should not be confused with the $r$ ``repeats'' that are internal to this
particular sorting algorithm $S$.

We illustrate this for our default sorting algorithm (using the best of $r=10$ k-means++ repeats)
in \fref{f:clips}(c), which shows 20 independent samples of $f^\tbox{rerun}_k$
with quantiles and means.
This shows that the first four spike types are stable
(as expected from the large waveform norms in panel (a) and well-separated
clusters in (b)), while the last four are less so.
The last ($k=8$) is the least stable, and from its tiny waveform it might be
judged not to represent a single neural unit.

\begin{rmk}
  We argue that 20 samples of any metric $f_k$ are more than
  sufficient for validation,
because the resulting standard deviation of the estimator of the mean is
  then somewhat
  smaller than the width of the underlying distribution of $f_k$.
  If a neuron has a wide distribution of $f_k$, it is unstable, and
  high accuracy in estimating $\bar{f}_k$ is not needed.
  This follows wisdom from the Monte Carlo literature
  \cite[Sec.~7.5]{mackayerice}.
\end{rmk}

However, as \fref{f:rerun_rconv} shows,
as a larger number of repeats $r$ is used,
the stability for each label $k=1,\dots,K$ grows.
By $r=100$ all stabilities have converged to very close to 1,
ie there is very little variation in labeling due to random initialization.
This is %simply
because picking the best of many repeats
finds a sum of squares distances to centroids that is
closer to the global minimum.
The algorithm has effectively become deterministic,
%(the global optimum for k-means is independent of initialization),
hence the rerun metric is uninformative.
%Whether or not this is a meaningful clustering, 
Since we wish to validate deterministic spike sorting
methods (eg, ones based on hierarchical or density-based clustering
\cite[Ch.~14--15]{zakibook}),
it is clear that one must introduce variation into the input $X$.
We now turn to such methods.

%It is standard in the clustering literature to use subsampling
%to assess stability \cite[sect.~17.3.1]{zakibook},
%ie, re-running $S$ on a smaller subsets of clips.

\bfi % ffffffffffffffffffffffffffffffffffffffffffffffffffffffffffffffffffffffffff
\mbox{%m
\begin{minipage}{2in}(a)\\
\ig{width=2in}{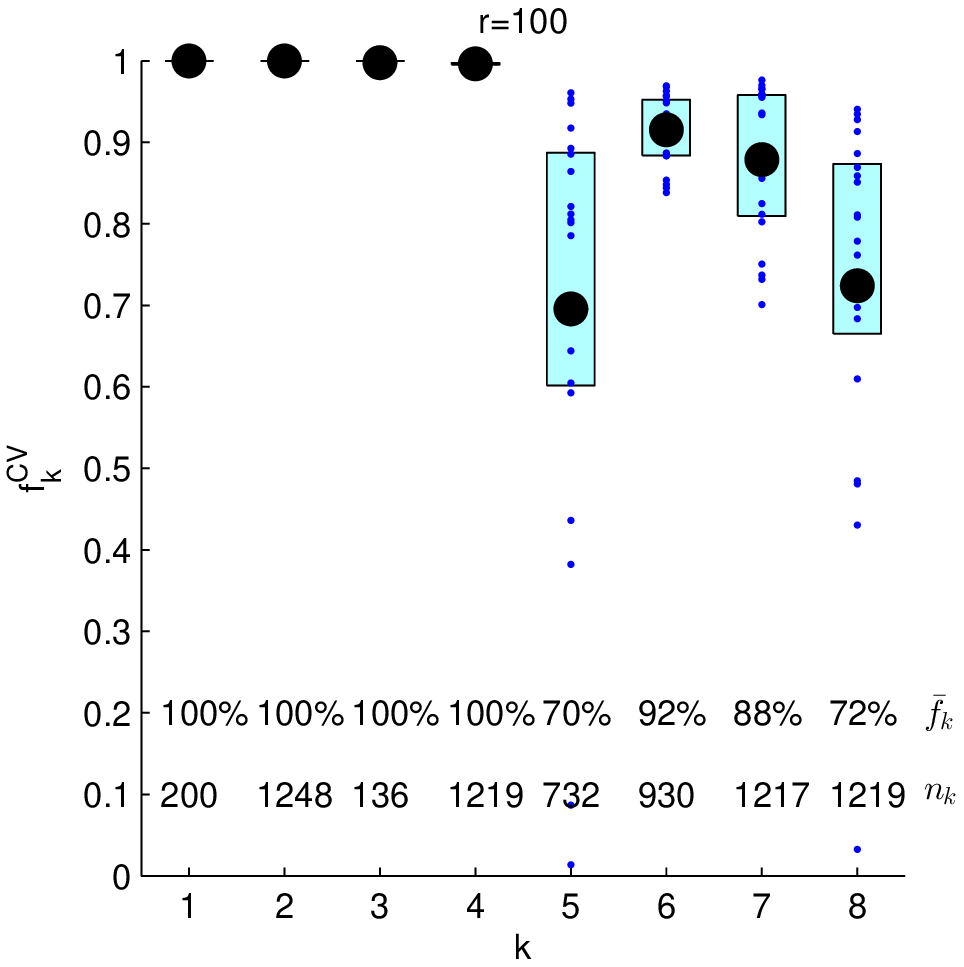}
\end{minipage}
  \begin{minipage}{1.4in}
\hfill(c)

\vspace{3ex}

(b)\\
\ig{width=1.4in}{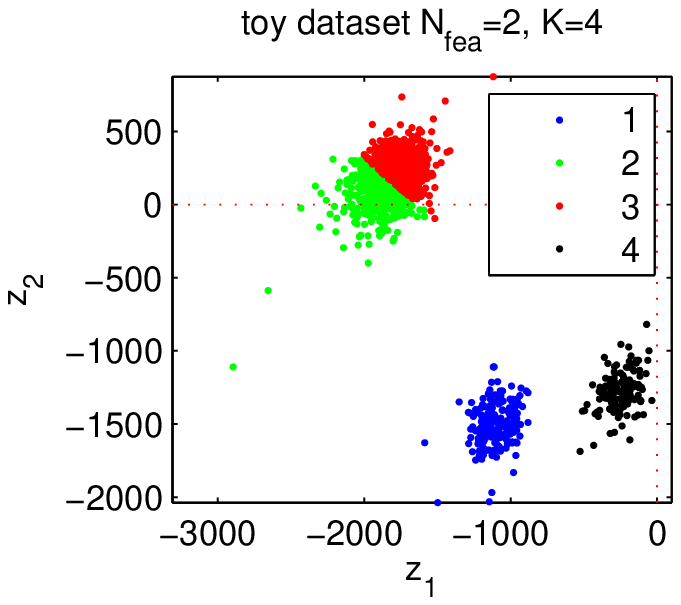}
\end{minipage}
  \begin{minipage}{1.1in}\ig{height=2in}{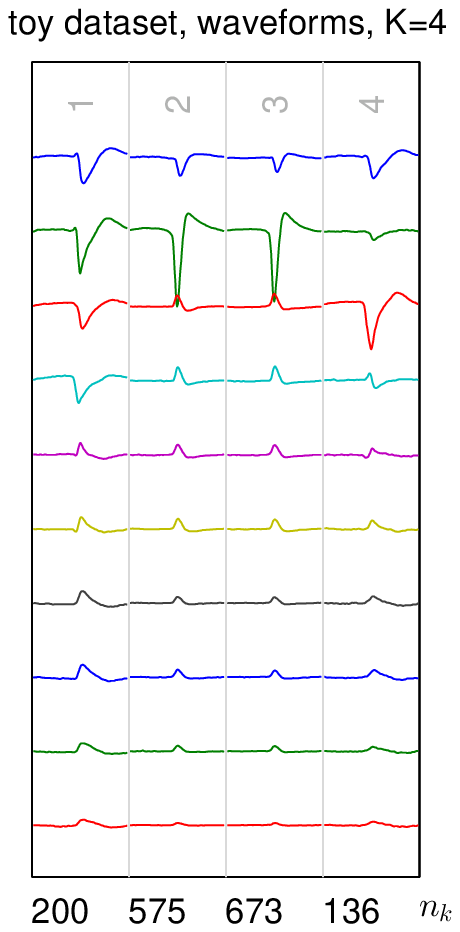}
\end{minipage}
  \begin{minipage}{1in}(d)\\
\ig{height=1.8in}{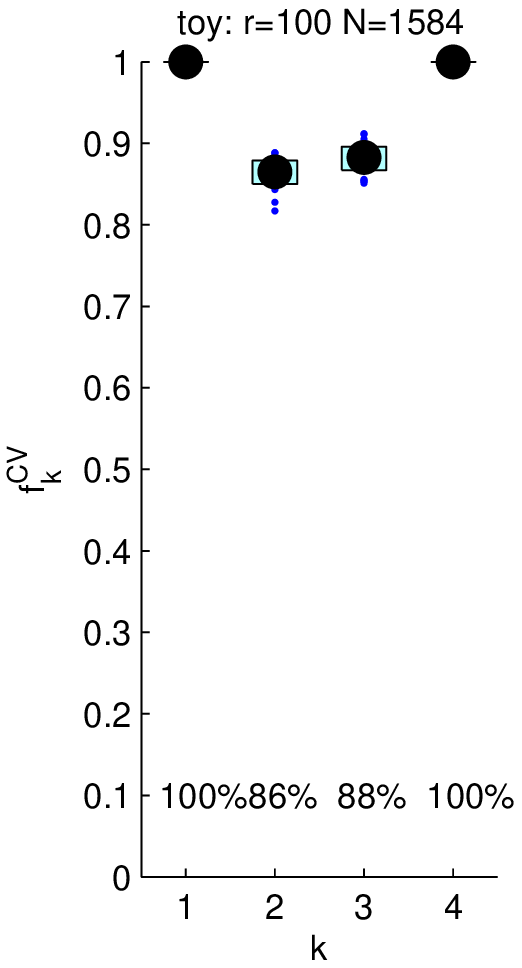}
\end{minipage}
  \begin{minipage}{1in}(e)\\
\ig{height=1.8in}{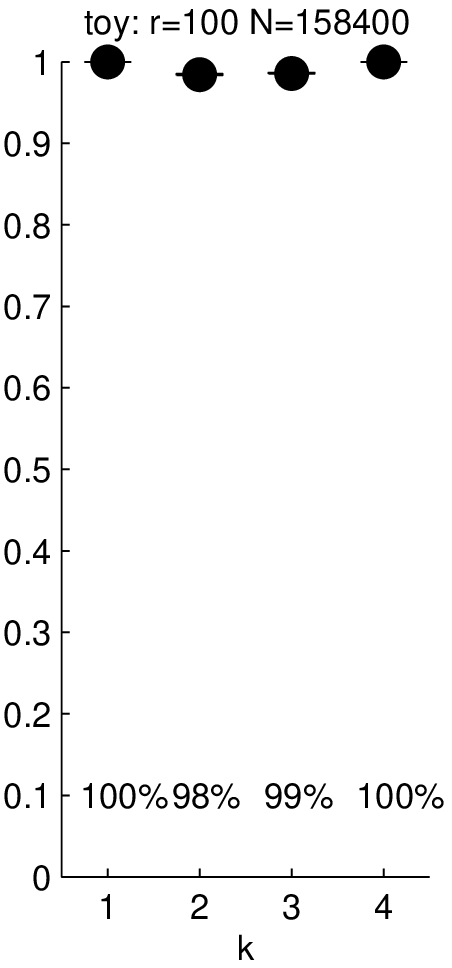}
\end{minipage}
}%m
\ca{Three-way cross-validation stability metric for PCA/k-means++ (best of $r=100$ repeats).
  (a) Stabilities for $K=8$, and populations from a single
  run of the sorter.
  (b) The two PCA components of the ``toy'' data
  (extracted from the labels $k\le 3$) with $N=1584$,
  labeled with PCA/k-means++ with $K=4$ and $\Nfea=2$.
  (c) Estimated waveforms from the run shown in (b).
  (d) Stability metrics for the $K=4$ labels on the data from (b).
(e) Same for an expanded dataset with a 100 times larger $N$.
}{f:cv}
\efi

% 33333333333333333333333333333333333333333333333333333333333333333333333333333
\subsection{A cross-validation metric using subsampling}
\label{s:cv}

In machine learning it is common to use cross-validation (CV) to
assess %the generalizability of
an algorithm.
The simplest version splits the data into training and test subsets,
and uses the performance on the test set to estimate the generalizability of
a classifier built using the training set.
Although CV requires an observed dependent variable, a similar idea can
be applied to spike sorting evaluation even in the absence of ground truth
information.
We divide the $N$ clips randomly into three similar-sized subsets I, II, and III,
and train a classifier $C_\tbox{I}$ with only the data I,
and a classifier $C_\tbox{II}$ with only the data II.
Finally we compute the best-permuted confusion matrix \eqref{Qbest}
between the labels produced by classifying (labeling)
the data III using $C_\tbox{I}$ and
using $C_\tbox{II}$,
then from this compute stabilities $f^\tbox{CV}_k$ for each spike type,
as before using \eqref{fk}.
Other more elaborate variants involving $m$-fold CV are clearly possible;
we do not test them here.
% why 3 sets? 2 is ok but has asymmetry between running C_I and S(II)

We now describe the classifier.
Our clip-based interface \fref{f:inter}(a) does not include
a classifier, yet we can build a simple one by taking the label of
the mean waveform \eqref{Wk} that is closest (in some metric) to each clip
in question.
That is, given a set of clips $X$ which gives spike-sorted labels $\kk = S(X)$,
the estimated waveforms $W^{(k)}$, $k=1,\dots, K$ are computed via \eqref{Wk};
they form the information needed for the classifier.
Applying the classifier to new clips $X' := \{x'_{mtj}\}$
gives new labels according to
$$
k'_j = \argmin_{k\in\{1,\dots,K\}} \sum_{mt} (x'_{mtj} - W^{(k)}_{mt})^2~,
$$
where for simplicity the $l_2$ norm has been chosen as the distance metric%
\footnote{This is appropriate for an iid Gaussian noise model; a more realistic
noise model is described in App.~\ref{a:large}. Also see \cite{prentice}.}.
Finally,
since the waveform norm ordering of \sref{s:ssalg} may vary for independent
subsamplings, then in order
to collect meaningful per-label statistics of $f^\tbox{CV}_k$
we must permute the labelings from later repetitions to best match the first.
For this we apply the Hungarian algorithm to the matrix of squared
distances ($l_2$-norms) between all pairs of mean waveforms in the
repetitions in question.
%
% ugh, was messier than I first thought. give formula?

\fref{f:cv}(a) shows this stability metric (for 20 independent 3-way splits)
for the PCA/k-means++ spike sorter with $K=8$ and $r=100$.
Notice that the last four spike types have stabilities in the range
70--90\%, in contrast to \fref{f:rerun_rconv}(c) for which they were all 100\%.
This shows that the three-way data subsampling introduces variation
that detects unstable label assignments,
even for an essentially deterministic spike sorter.

We now show that the 3-way CV metric is able to identify
an erroneously split cluster.
% Is this too much human judgment?
%
For this we generate a simple ``toy'' dataset comprising just the $N=1584$ clips
with labels $k_j=1$, 2 or 3 from the default $K=8$ and $r=100$ spike sorting
(\fref{f:clips}).
Thus there are three
very well-separated true clusters in feature space with
a clear ground-truth; see \fref{f:cv}(b).
For this toy dataset we pick a standard sorting algorithm
of PCA/k-means++ with $K=4$ (chosen to be erroneously large),
$r=100$, and $\Nfea=2$
(to aid visualization).
Running this algorithm on the toy data gives the labeling
in two dimensions of feature space shown in
\fref{f:cv}(b), and waveforms of \fref{f:cv}(c):
the largest true cluster is erroneously split into two nearly-equal pieces,
which are given new labels $k=2,3$ (hence the similar pair of waveforms in (c)).
There are many ways to summarize accuracy\footnote{%
All four clusters have precision 1, but $k=2,3$ have recall around 0.5.
}
here \cite[Sec.~17.3.1]{zakibook}:
filling the confusion matrix \eqref{Qbest} of size $3\times 4$ between
truth and the $K=4$ clustering gives the F-measures \eqref{fk}
for the three true labels as $f_2=0.701$, $f_1=f_3=1$.
Running the CV metric on this $K=4$ sorting algorithm
results in the stabilities in \fref{f:cv}(d):
a mean around 87\% for $k=2,3$ (somewhat higher than measures of accuracy),
and 100\% for the
other (unsplit) clusters.
The point of our framework is that this conclusion is
reached via {\em only} the standard interface $S$, ie
without access to the feature vectors shown in (b), nor of course
the ground truth.

However, the variation in labeling
induced by subsampling is small compared to the known accuracy
(one reason being that the small $\Nfea$ makes for high stability),
and furthermore
dies away as the number of clips grows, $N\to\infty$.
We illustrate this in \fref{f:cv}(e), which shows the 3-way CV metric
for the same algorithm as (d) but running on a set of clips 100 times
larger, generated as described in App.~\ref{a:large}.
The stabilities of the $k=2,3$ labels of the split cluster are now around 98.5\%,
hardly an indication of instability.
%ie 10 times closer to perfect stability than before.
This is consistent with a central limit theorem proven \cite{shamir}
for stability metrics in a wide class of clustering algorithms,
when their global optimum is unique,
implying in our setting
that $$1 - f^\tbox{CV}_k = \bigO(N^{-1/2}), \qquad \mbox{ for each $k$} ~.$$
Indeed, increasing $N$ by a factor of 100 decreased the deviation of $f_k$
from unity by around a factor of 10.
We expect the prefactor for this deviation to be smaller
when the split cluster is more {\em non-spherical}:
resampling induces high instability in splitting a spherical cluster,
but in k-means a non-spherical cluster (as in \fref{f:cv}(b)) tends to
split consistently along a hyperplane perpendicular to its longest axis.
%
% explain more why k-means more stable if higher aspect ratio blob?

This unfortunate
phenomenon that, regardless of the apparent quality of the labeling,
stability tends to unity as $N$ grows has been known in the clustering literature
for some time \cite{krieger}.
% discuss fixes:
Renormalizing a global stability metric by $\sqrt{N}$
is possible \cite[Sec.~3.1.1]{luxburg},
but it is not clear how to do this when a {\em per-neuron} metric is
needed.
% todo: work on this!
%
On the other hand, restricting to small-$N$ subsamples to induce instability
has the major disadvantage that
the spike sorter is then never validated with any dataset of the desired size.

Since we seek a scheme which can validate spike sorting applied to
arbitrarily large datasets,
we must move beyond subsampling, and explore perturbing the clips themselves.

%A standard idea from statistics is
%to introduce variation by running the
%algorithm repeatedly on random samples of size smaller than $N$
%taken without replacement
%(subsampling \cite[sect.~17.3.1]{zakibook}),
%or samples of size $N$ taken with replacement (bootstrapping \cite{luxburg}).
%The disadvantage of the former is that it does not assess spike sorting
%performance on datasets of the desired size $N$
%-- this will affect density-based clustering algorithms
%if the distance parameter is not adjusted.
%the latter also
%has complications due to weighting repeated points \cite{luxburg}.
%Again, as the sample sizes grow, stability is expected
%to converge to 1 and hence cease to be a useful measure.
%this has been appreciated in the clustering context

\bfi % fffffffffffffffffffffffffffffffffffffffffffffffffffffffffffffffffffff
\mbox{}\hspace{.3in}\ig{width=6.1in}{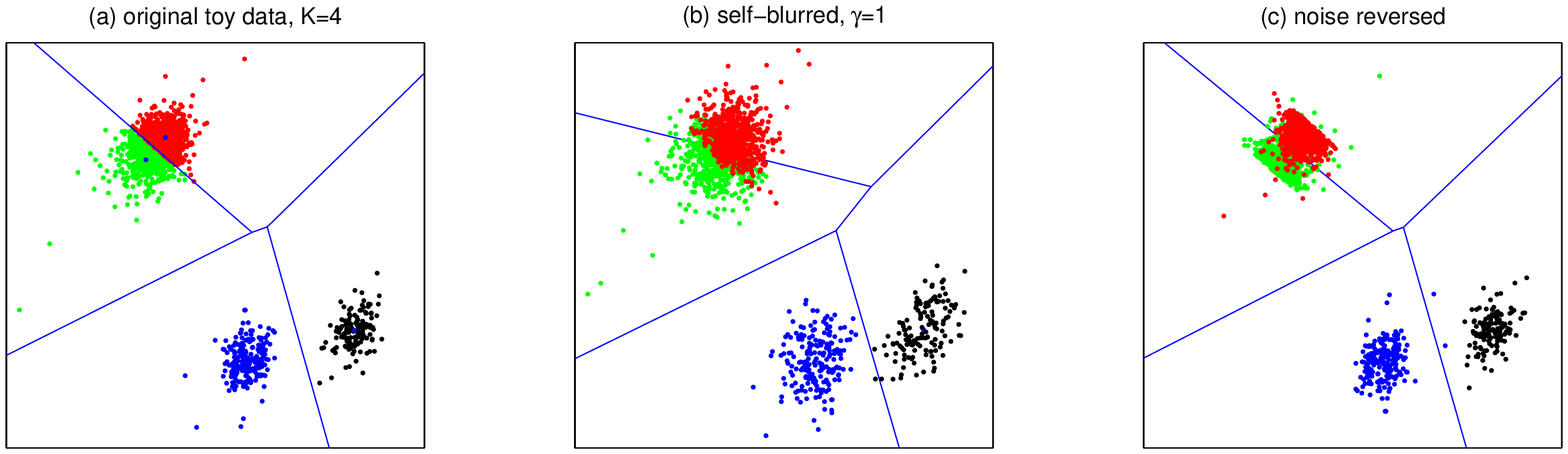}
\\
\raisebox{-1in}{\ig{height=2in}{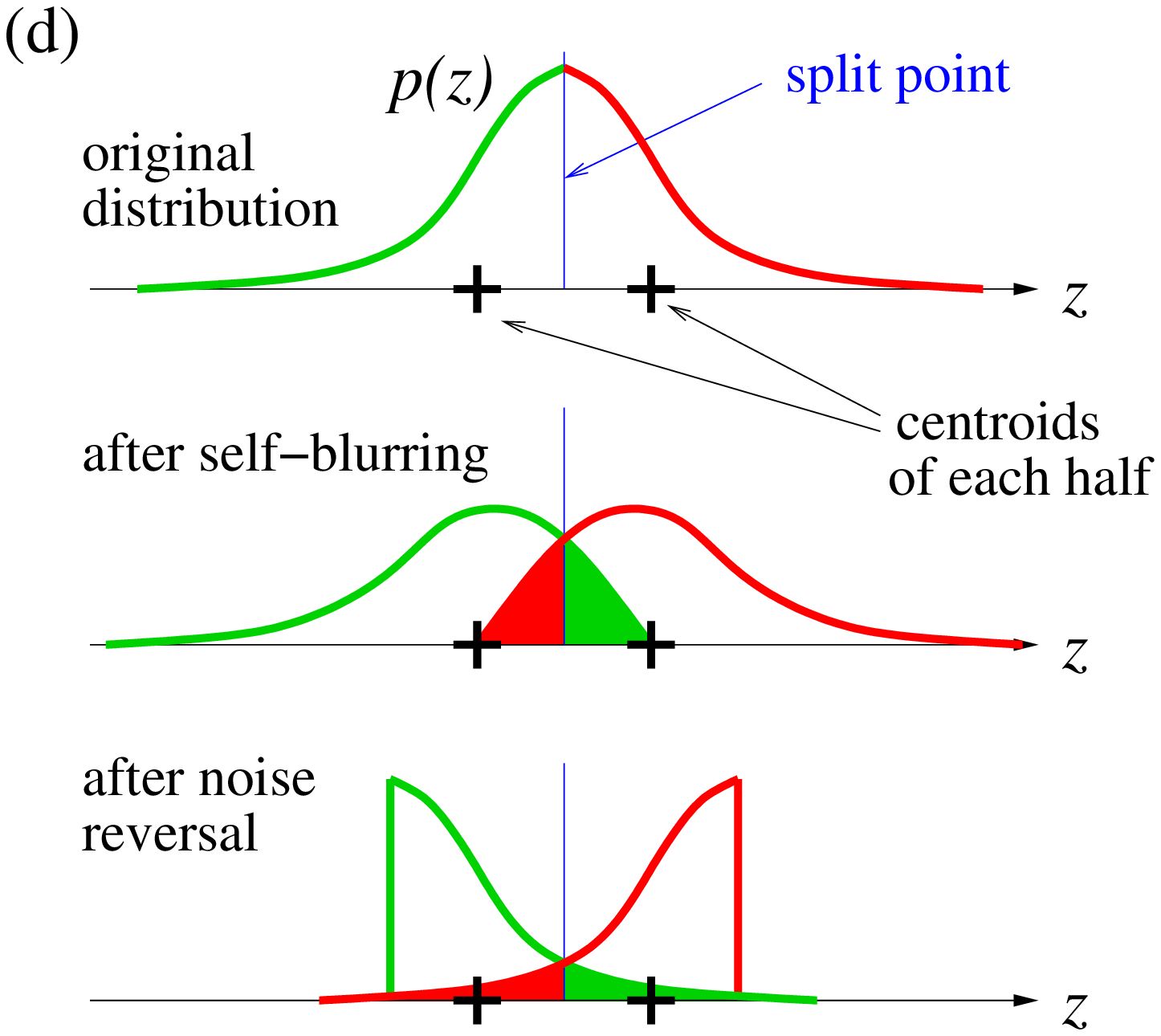}}
\;
\begin{minipage}{4.2in}
(e) \hspace{1.9in} (f)
\\
\ig{height=1.9in}{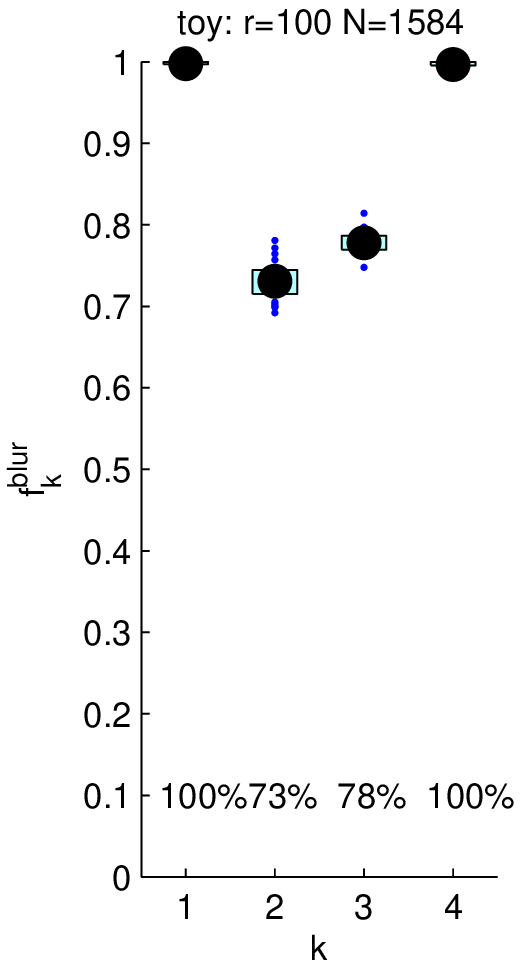}
\ig{height=1.9in}{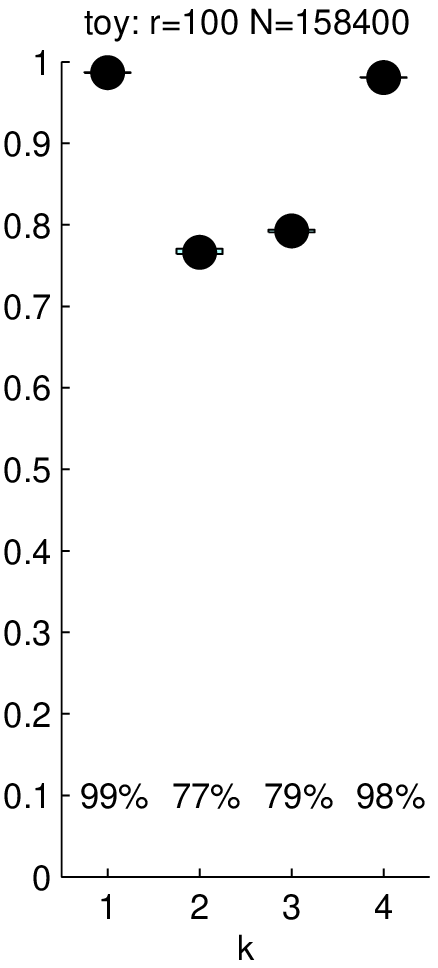}
\;
\ig{height=1.9in}{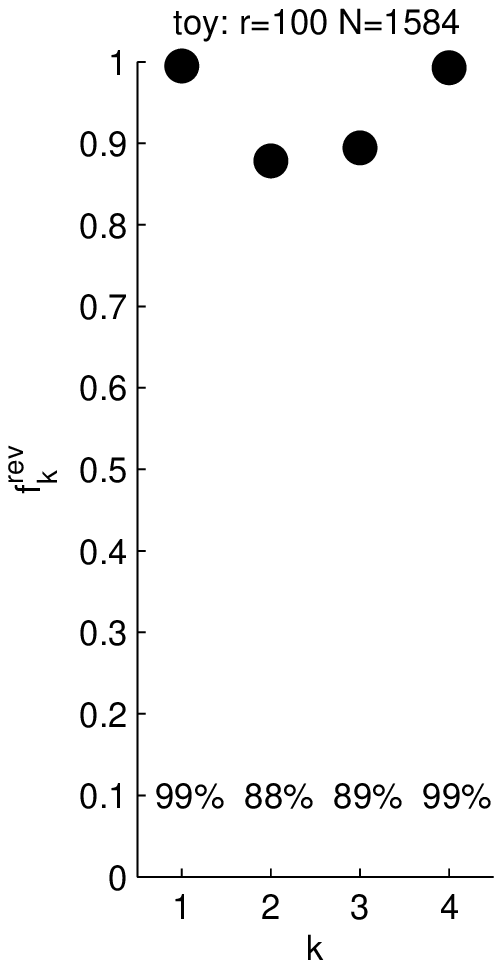}
\ig{height=1.9in}{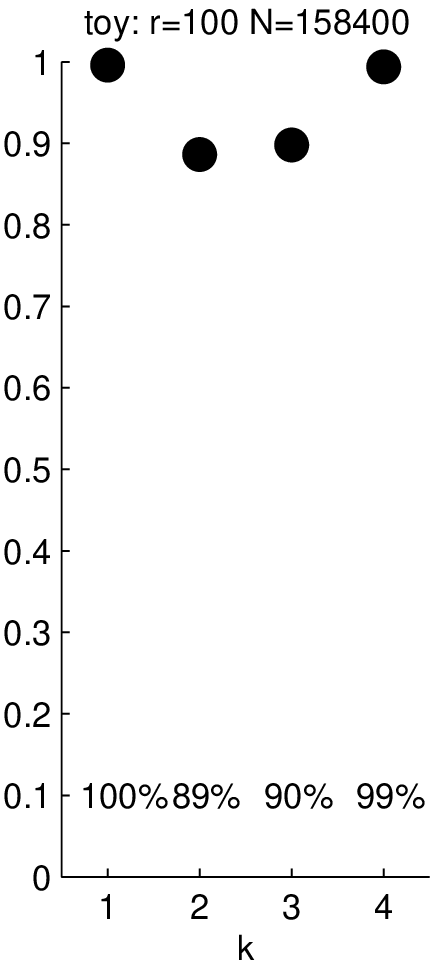}
\end{minipage}
\ca{Data perturbation validation metrics illustrated on the
``toy'' clips (from labels $k\le3$), sorted with
PCA/k-means++ with $K=4$, $r=100$, $\Nfea=2$.
(a)--(c) shows the 2D PCA feature space.
(a) Original points labeled by color as in \fref{f:cv}(b),
showing k-means centroids and decision boundaries.
(b) Points after self-blurring and (c) after noise reversal;
colors show the {\em original} labelings, to highlight movement,
and new decision boundaries.
(d) Idealized 1D picture: only the shaded parts of the distribution change
labeling.
Stability metrics are shown for
(e) self-blurring and (f) noise reversal,
for the original toy data size $N$, and a 100 times larger $N$.%
}{f:pert}
\efi

% PPPPPPPPPPPPPPPPPPPPPPPPPPPPPPPPPPPPPPPPPPPPPPPPPPPPPPPPPPPPPPPPPPPPPPPPPPP
\subsection{Metrics based on data perturbations without subsampling}
%self-blurring and noise reversal}
\label{s:pert}

\fref{f:cv}(b)--(e) shows that for large $N$, stability under
subsampling fails to indicate a highly erroneous split cluster.
By examining feature space (\fref{f:cv}(b))
it is clear that a decision boundary of k-means passes through
a high-density region.
Can one create a metric that quantifies this
undesirable property without access to feature space,
using only the standard interface?
We now present two such metrics, both of which involve perturbing the data.

% rmk re could just add noise but don't know noise model

\subsubsection{Self-blurring}
\label{s:blur}

Firstly we describe a method that we name {\em self-blurring} for reasons
that will become clear.
Let $\kk = S(X)$ be the output of the sorter on the full set of clips,
and $\{W^{(k)}\}_{k=1,\dots,K}$ be the mean waveforms found via \eqref{Wk}.
Let $\pi$ be a permutation of the $N$ clips that randomizes {\em only
within each label class}.
Precisely, if $J_k:= \{j:k_j=k\}$ is the index set of the $k$th label type,
and $\pi_k$ is a random permutation of $n_k$ elements,
then the overall permutation $\pi$ is defined by
$$
\pi(J_k(i)) \;=\; J_k(\pi_k(i)), \qquad i=1,\dots,n_k
$$
holding for each label type $k=1,\dots,K$.
From this we create a perturbed clip dataset $\tilde{X}$ with elements
\be
\tilde{x}_{mtj} \;=\; x_{mtj} + \gamma (x_{mt,\pi(j)} - W^{(k_j)}_{mt}),
\qquad m=1,\dots,M, \; t=1,\dots,T,\; j=1,\dots,N,
\label{blur}
\ee
where $\gamma>0$ is a parameter controlling the size of the perturbation.
Running the sorter on this new data gives $\tilde{\kk} = S(\tilde X)$,
then we compute the best-permuted confusion matrix \eqref{Qbest}
between labels $\kk$ and $\tilde\kk$, and from it obtain
the per-label stabilities
$f_k^\tbox{blur}$ via \eqref{fk}.

The bracketed expression in \eqref{blur} can be interpreted as a random
sample from the ``noise distribution'' of the $k_j$th cluster
in clip signal space, relative to its mean waveform (centroid).
So, with $\gamma=1$, %(our default choice),
the cluster distribution about its centroid is convolved (blurred)
with itself.
The result is that, even if a decision boundary remains fixed,
if there is density near this boundary
then points are carried across it by the convolution,
thus change label, and are registered as instability.
The new data $\tilde X$ is consistent with the original $X$
in the sense that it differs only by additive noise of exactly
the type observed within each cluster.
No assumption on the form of the noise model or noise level is made.

Since $\RR^{MT}$ (signal space) is hard to visualize, we again
``peek under the hood'' of our PCA/k-means++ algorithm, by in
\fref{f:pert}(b) showing the result of self-blurring on the 2D feature vector
points. Notice that there is significant overlap of red and green points,
and rotation of one of the decision lines.
The self-convolution also results in larger clusters,
which may cause label exchange between clusters that were distinct
(eg see the tails of the lower two clusters).
For an isolated Gaussian cluster, its size (noise level)
grows by a factor $\sqrt 2$.
\fref{f:pert}(e) shows the distribution of $f_k^\tbox{blur}$ for
20 samples of $\tilde X$.
The mean around 75\% for $k=2,3$,
being similar to the actual F-measure accuracy of 70\%,
is a clear indicator of instability.
For the
dataset of size $100 N$ this also holds;
thus, we have cured the large-$N$ vanishing of instability present with CV.

%we take each clip, perturb it in a random manner consistent with
%the empirical noise, and measure stability of the sorter on the full set of

We show an idealized example where points are drawn from a
symmetric distribution $p(z)$
in a 1D variable $z$, erroneously split into two clusters, 
in the top two pictures in \fref{f:pert}(d). The decision boundary
is unchanged by self-blurring. The mass of
each half of the distribution that changes label is shaded;
this controls the size of $1-f_1^\tbox{blur}$.
If this distribution were a Gaussian, and $\gamma=1$,
we can solve analytically (see App.~\ref{a:gauss}) that
$f_1^\tbox{blur} = 1-[\erf((2\pi)^{-1/2})]^2 \approx 0.817$.
The same holds for
a multivariate Gaussian in signal space split along a plane of symmetry.
The reported stabilities in \fref{f:pert}(e)
also are similar to this value, even though
clusters are not expected to be Gaussian \cite{fee1996}.
%(even though we do not know if a Gaussian is a good model
%for the distribution around centroids in our toy
%dataset).

Later we will vary $\gamma$:
a smaller $\gamma$ tends to increase all reported stabilities,
but also causes less inflation of clusters (ie less increase in noise level),
which may better reflect an algorithm's stability
for close, but distinct, clusters.
%varying $\gamma$ may be a useful statistical tool that we leave for future work.

\subsubsection{Noise reversal}
\label{s:rev}

Our second perturbation method has a similar flavor: using the
above idea of noise (deviation) relative to a cluster centroid, we simply
negate the noise. That is, we replace \eqref{blur} by
\be
\tilde{x}_{mtj} \;=\; 2W^{(k_j)}_{mt} - x_{mtj}~,
\qquad m=1,\dots,M, \; t=1,\dots,T,\; j=1,\dots,N,
\label{rev}
\ee
and proceed exactly as above to get $f_k^\tbox{rev}$.
Since this is a deterministic procedure, there is no need to resample;
only two spike sorter runs are needed (the original and the noise-reversed).
This is shown in \fref{f:pert}(c), and lower picture of (d):
each cluster is inverted about its centroid.
Label exchange occurs in a split cluster because the tails of the distributions
are made to point inwards, and bleed across the boundary.
The results for this toy dataset appear in \fref{f:pert}(e):
the erroneously-split labeling induces around 89\% stability
for both the original $N$ and a 100 times larger $N$.
This is more stable than both self-blurring and the true accuracy---a
disadvantage---%
however its advantages are that the clusters are not artificially enlarged,
that there are no free parameters, and no resampling.
%
% also, disadv is that non-inversion-symmetric noise gets changed
% in a non-physical way.  Ie weird overlaps of strange shaped clusters.

In the idealized 1D Gaussian case, and the multivariate Gaussian split along a symmetry plane, analytically (see App.~\ref{a:gauss})
we get
$f_1^\tbox{rev} = \erf(2/\sqrt{\pi}) \approx 0.889$,
which is very close to the observed values, and explains
why stabilities are expected to be closer to 1 than for self-blurring
with $\gamma=1$.

\bfi % fffffffffffffffffffffffffffffffffffffffffffffffffffffffffffffffffffff
\begin{minipage}{2in}
(a)
\\ 
\ig{width=2in}{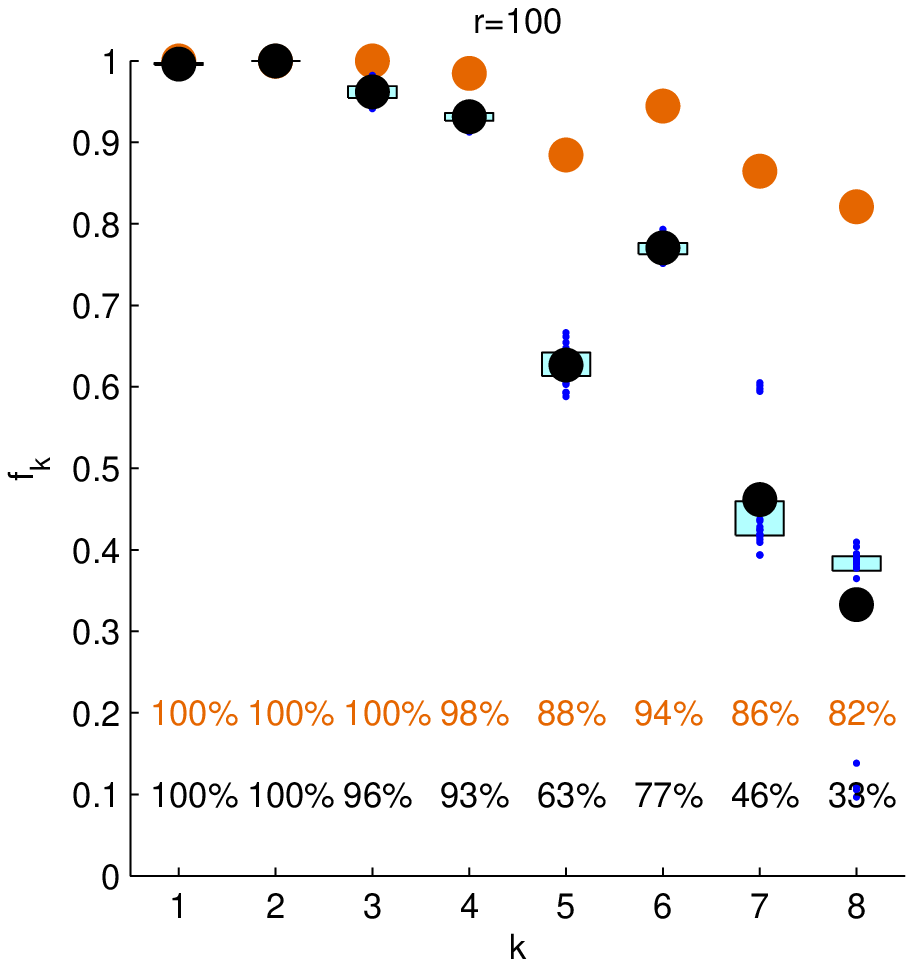}
\end{minipage}
\begin{minipage}{2.4in}
  (b)
  \\
  \raisebox{-1in}{\ig{width=2.4in}{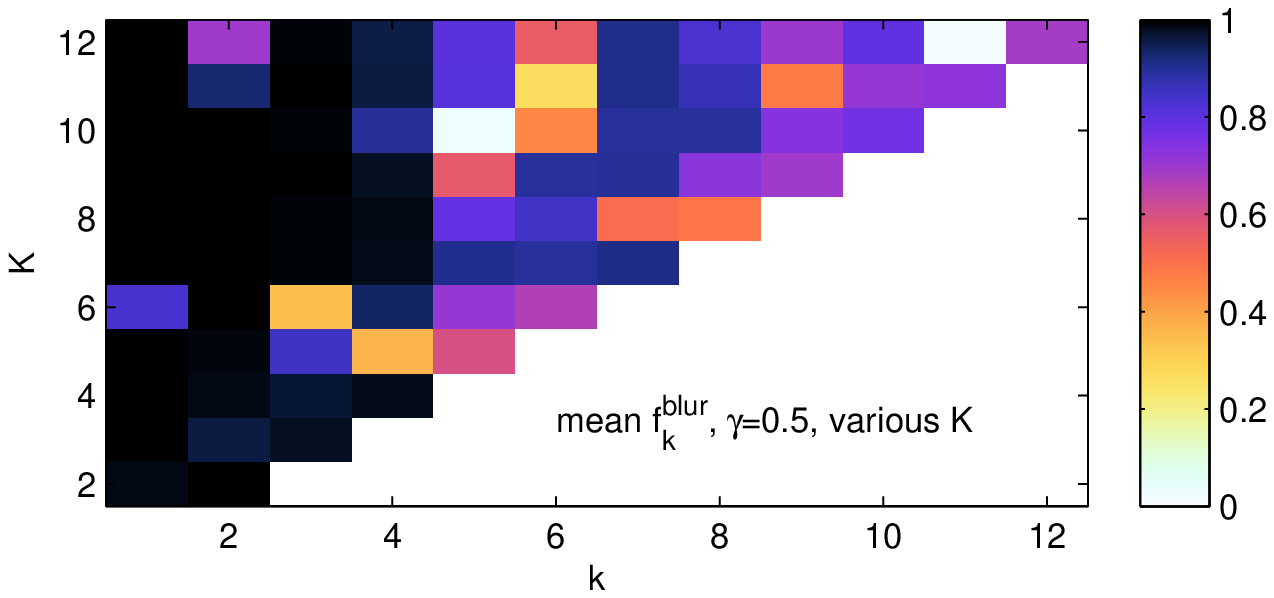}}\\
  (c)
  \\
  \raisebox{-1in}{\ig{width=2.4in}{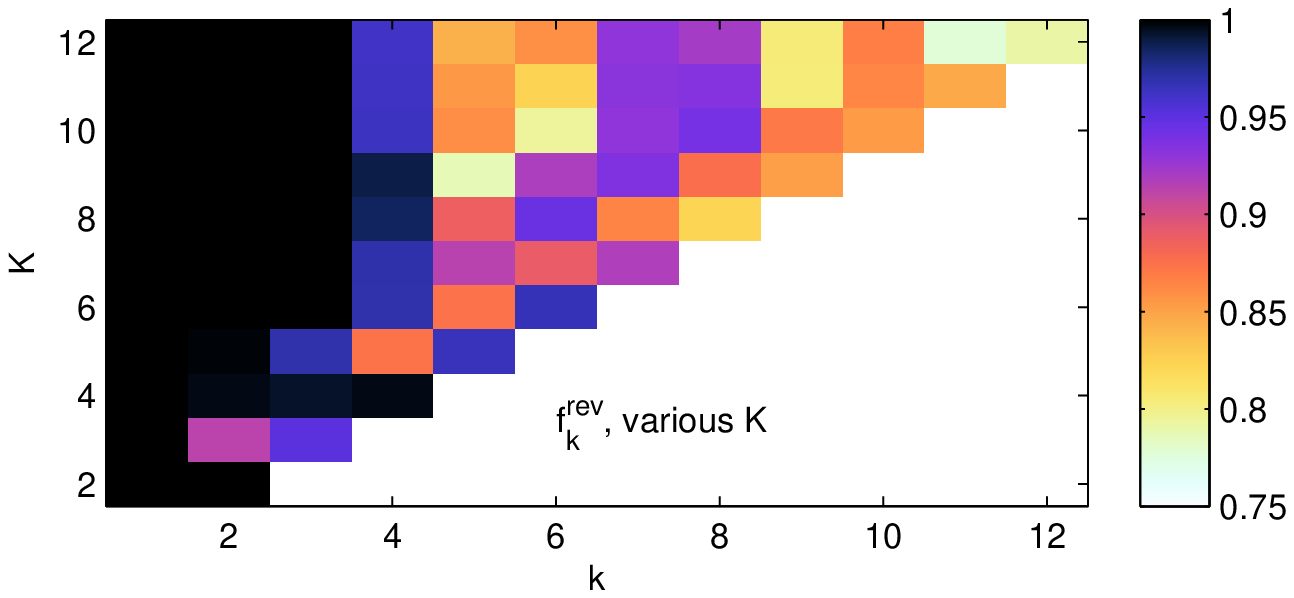}}
\end{minipage}
\begin{minipage}{2in}
  (d)\raisebox{-1in}{\ig{width=2in}{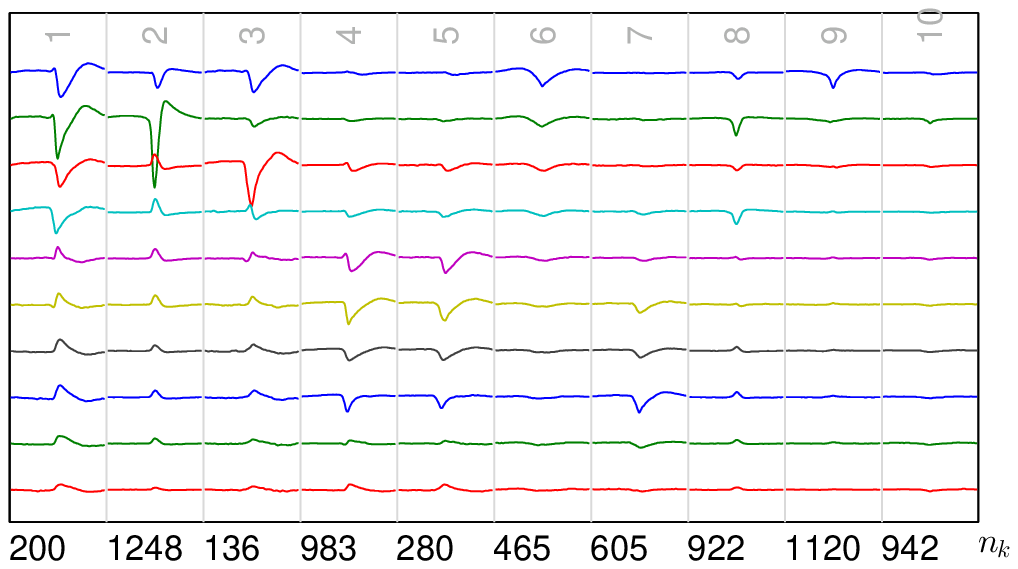}}
  \\
  (e)\raisebox{-1.2in}{\ig{width=2in}{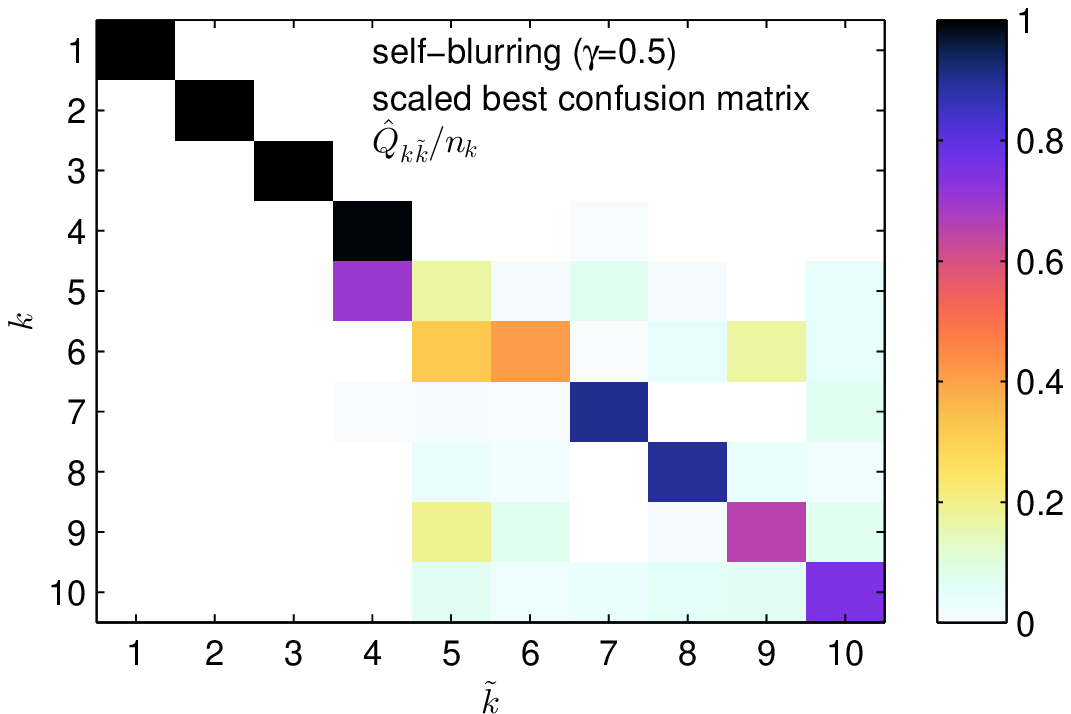}}
\end{minipage}
\ca{
  Perturbation metrics for default clips dataset with PCA/k-means++ with
  $r=100$.
  (a) self-blurring ($\gamma=1$, shown in black) and noise reversal
  (shown in orange) metrics for $K=8$.
  (b) self-blurring ($\gamma=0.5$) metric $\bar{f}_k^\tbox{blur}$
  for various $K=2,\dots,12$
  (each row a different $K$).
  (c) same for noise reversal metric $f_k^\tbox{rev}$.
  (d) mean waveforms for $K=10$, and (e) self-blurring scaled best confusion
  matrix $\hat{Q}_{k,\tilde{k}}/n_k$.
  }{f:pertclips}
\efi

\subsubsection{Perturbation metric tests for the default algorithm}
\label{s:perttest}

We return to the full clips data
from \sref{s:ssalg} with PCA/k-means++ with the default $\Nfea=10$
and $r=100$.
\fref{f:pertclips}(a) compares the self-blurring and noise reversal metrics
for $K=8$:
we see that noise reversal has all stabilities roughly 3 times closer to
unity than self-blurring at $\gamma=1$, but otherwise similar
relative stabilities vs neuron label $k$.
%Both show similar relative stabilities as CV in \fref{f:cv}(a).

Since we cannot presume that $K=8$ is optimal for
extracting the largest number of stable neurons,
in \fref{f:pertclips}(b)--(c) we show the mean stabilities varying
$K=2,\dots,12$.
Both show that the first four neurons are stable for most $K$ values,
and show that one or two others are also quite stable for $K\ge 8$
(these are neurons $k=7,8$ for $K\ge 10$).
Notice that, for noise reversal, no stabilities drop below 0.75,
ie the stability range is compressed towards unity.

We study $K=10$ in more detail, since for it, neuron $k=5$ shows
a self-blurring stability close to zero.
\fref{f:pertclips}(d) shows the mean waveforms for this $K$;
indeed $k=4,5$ have very similar waveforms so would probably
be deemed erroneously split in practice.
The (best-permuted) confusion matrix $\hat{Q}$ is a useful
diagnostic:
as the vertical ``domino'' in \fref{f:pertclips}(e) shows,
$k=4$ and 5 have been merged by self-blurring.
Off-diagonal entries such as $\hat{Q}_{6,9}$ show that $k=6,9$ are
prone to mixing, which is borne out by their visually similar
waveforms.
%while the horizontal domino shows $k=6$ has been split.
% by the noise addition,
The point is that our metrics quantify such judgments,
%in a noise-model agnostic way,
without a human having to examine whether waveforms are ``close''
given the particular noise (and firing variation) of the experiment.
%The smallest (``noise cluster'') waveform $k=10$ mixes with all the
%other smaller amplitude waveforms $k=5,\dots,9$, reducing
%their stability, although 7 and 8 have the least off-diagonal mixing.

Finally, we observe that for $K=4$, all supposed neurons are very stable
(under both metrics), even though, with our knowledge of the larger
number of distinct waveforms extracted for greater $K$, it is obviously
not accurate.
This illustrates that {\em stability is necessary but not sufficient for
accuracy}.
%One cannot presume accuracy for stable neurons,
%but inaccuracy can be deduced for supposed neurons that are unstable.
% cite Hennig clusterwise?

\begin{rmk}
Although we illustrated the self-blurring and noise reversal metrics
with a label assignment (clustering)
scheme involving hard decision boundaries (hyperplanes for k-means),
we propose a broader view:
since both metrics involve perturbing the clip
data in a manner consistent with the noise, any reproducible
neural unit output by a spike sorting algorithm
should be stable under these metrics.
\end{rmk}

% Accuracy section: ?
%do all 4 methods, with synthetic clips from given set of $W$ plus noise?
%accuracy $\le$ stability, but only roughly.
%extract clips from Quiroga synthetic time-series data?

% TTTTTTTTTTTTTTTTTTTTTTTTTTTTTTTTTTTTTTTTTTTTTTTTTTTTTTTTTTTTTTTTTTTTTTTTTTTTT
\section{Time-series based validation schemes}
\label{s:tseries}

%Building upon the above metrics,
We now turn to the second, more general, interface of \sref{s:inter},
where the spike sorter extracts both times $t_j$ and labels $k_j$ from a time
series $Y$ of multi-channel recorded data.
To compare two outputs of such a sorter,
say $\{t_j,k_j\}_{j=1}^{N_s}$ and $\{t'_j,k'_j\}_{j=1}^{N_s'}$,
where $k_j\in\{1,\dots,K\}$, $k'_j\in\{1,\dots,K'\}$, and $K$ and $K'$
are the two
(not necessarily equal) numbers of types of neurons found,
one must match spikes from the first run to those in the second.
A simple criterion is that matched firing times be
within $\epsilon$ of each other; we fix $\epsilon = 0.5 $ ms in our tests.
Specifically, let a matching $\mu$ be a one-to-one map%
\footnote{$\mu$ can also be expressed as a subgraph of the
  complete bipartite graph between $N_s$ nodes and $N'_s$ nodes,
  whose vertex degrees are at most one.}
from a subset of $\{1,\dots,N_s\}$
to an (equal-sized) subset of $\{1,\dots,N'_s\}$ such that
\be
|t_j - t'_{\mu(j)}| \le \epsilon
~, \qquad \mbox{for all $j$ matched in the map.}
\label{tdiff}
\ee
Given any such $\mu$, the confusion matrix is defined by
\be
Q^{(\mu)}_{k,k'} = \#\{j : \mu(j) \mbox{ exists}, \;k_j=k, \;k'_{\mu(j)} = k' \}
~,
\qquad \mbox{ for } 1\le k\le K, \; 1\le k' \le K'~.
\label{Qmu}
\ee
Since there may be unmatched spikes $j$ in the first run
(for which we write $\mu(j) = \ns$), and unmatched spikes $j'$ in the
second run (for which we write $\mu^{-1}(j') = \ns$),
a row and column is appended for these cases to $Q^{(\mu)}$, ie,
\bea
Q^{(\mu)}_{k,\ns} &=& \#\{j : \mu(j)=\ns, \;k_j=k \}
~,
\qquad \mbox{ for } 1\le k\le K~,
\\
Q^{(\mu)}_{\ns,k'} &=& \#\{j : \mu^{-1}(j')=\ns, \;k'_{j'}=k' \}
~,
\qquad \mbox{ for } 1\le k' \le K'~,
\label{Qmuns}
\eea
and set $Q^{(\mu)}_{\ns,\ns} = 0$.
The $(K+1)$-by-$(K'+1)$ matrix $Q^{(\mu)}$ defined by \eqref{Qmu}--\eqref{Qmuns}
we call the {\em extended confusion matrix}; see \fref{f:tseries}(a).

Since a sorting algorithm may arbitrarily permute the neuron labels,
as before we need to find the ``best'' confusion matrix.
However, now one must search
over permutations $\pi$ of the second label {\em and} over matchings $\mu$, ie
find
\be
\hat{Q}_{k,k'} := Q^{(\hat\mu)}_{k,\hat\pi(k')}~,
\qquad \mbox{ where } \quad
\{\hat\mu, \hat\pi\} = \argmax_{\mu\in{\cal M}, \pi\in S_{K'}} \sum_{k=1}^{\min(K,K')} Q^{(\mu)}_{k,\pi(k)}
~,
\label{Qextbest}
\ee
where ${\cal M}$ is the set of all maps $\mu$ satisfying \eqref{tdiff}.
In practice we find that a simple three-pass algorithm
performs well at typical firing rates (optimizing $\pi$
using a greedy-in-time-difference
algorithm for matching, then with $\pi$ fixed, optimizing $\mu$ via
two passes of greedy matching).
We do not know the complexity of finding the exact solution.

%we do not know if there is a polynomial time algorithm
%for the exact solution.
% could use Hungarian within bursts with dist set to infty for combos that
% are >eps apart in time.

\bfi  % fffffffffffffffffffffffffffffffffffffffffffffffffffffff
\ig{width=6.5in}{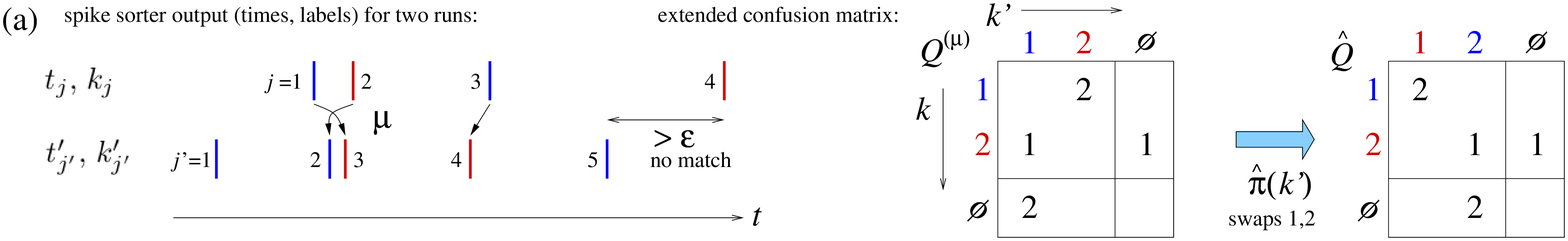}
\\
\vspace{1ex}
\mbox{}
\\
\begin{minipage}{2in}
  (b)\\
  \ig{width=2in}{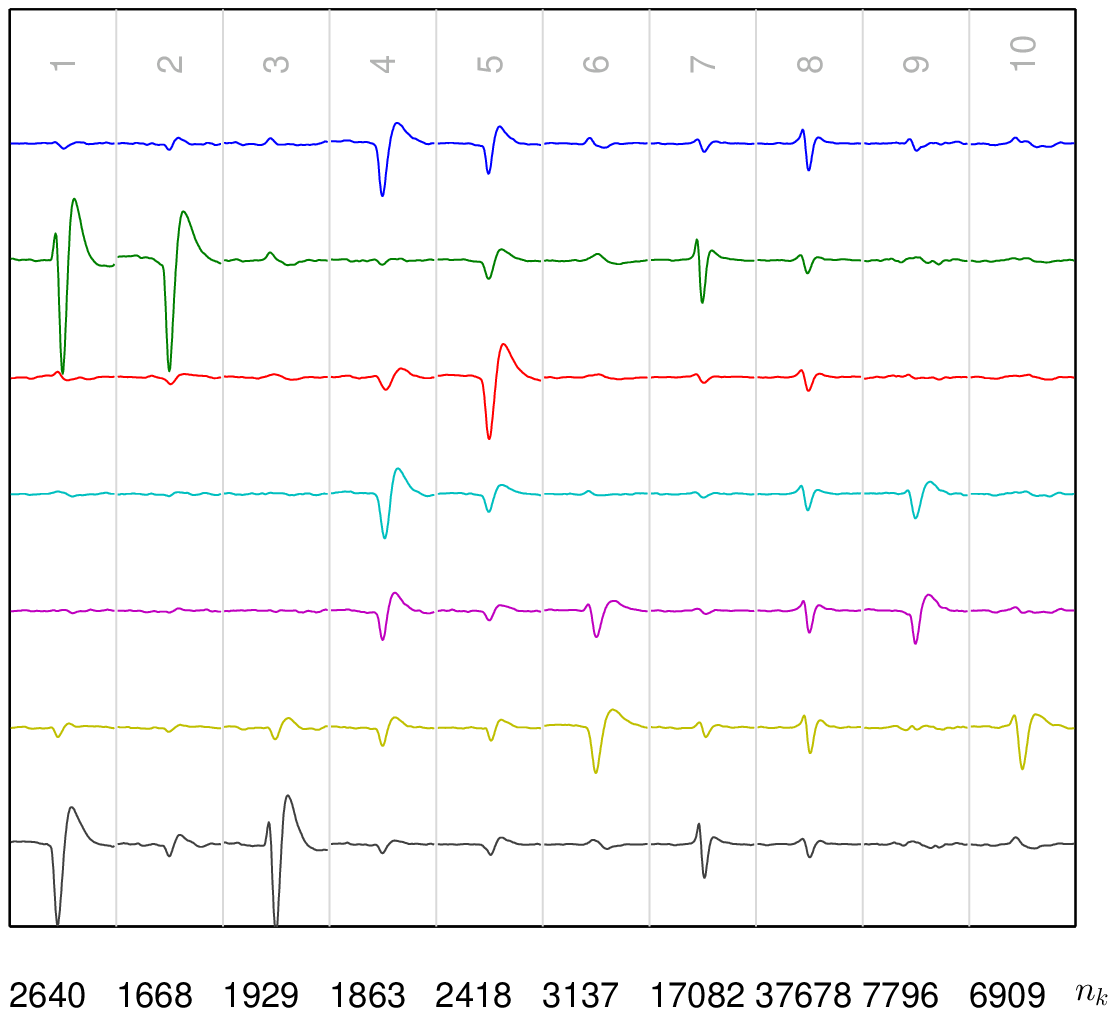}\\
  \\
  (c)\\
  \ig{width=2in}{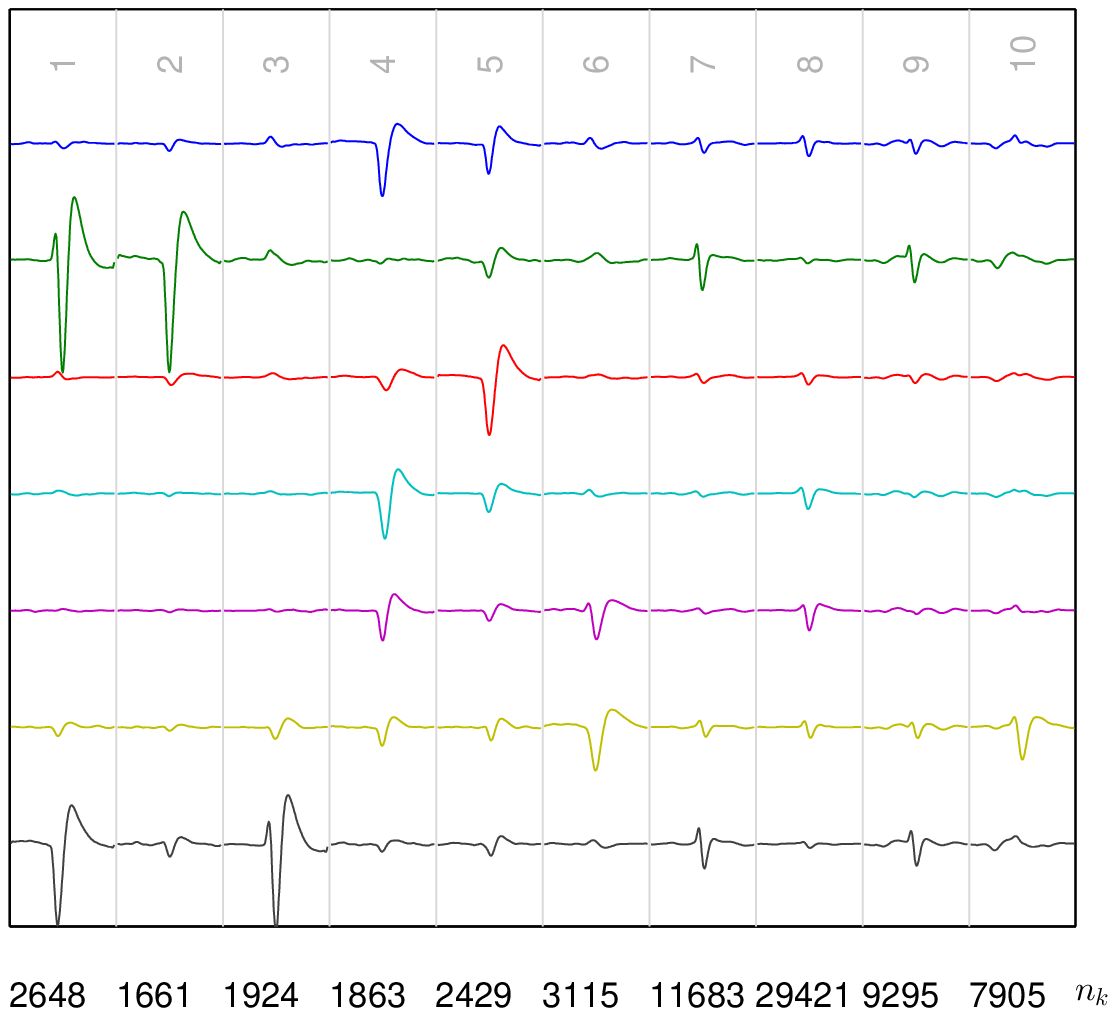}
\end{minipage}
\begin{minipage}{2.4in}(d)\\
  \ig{width=2.4in}{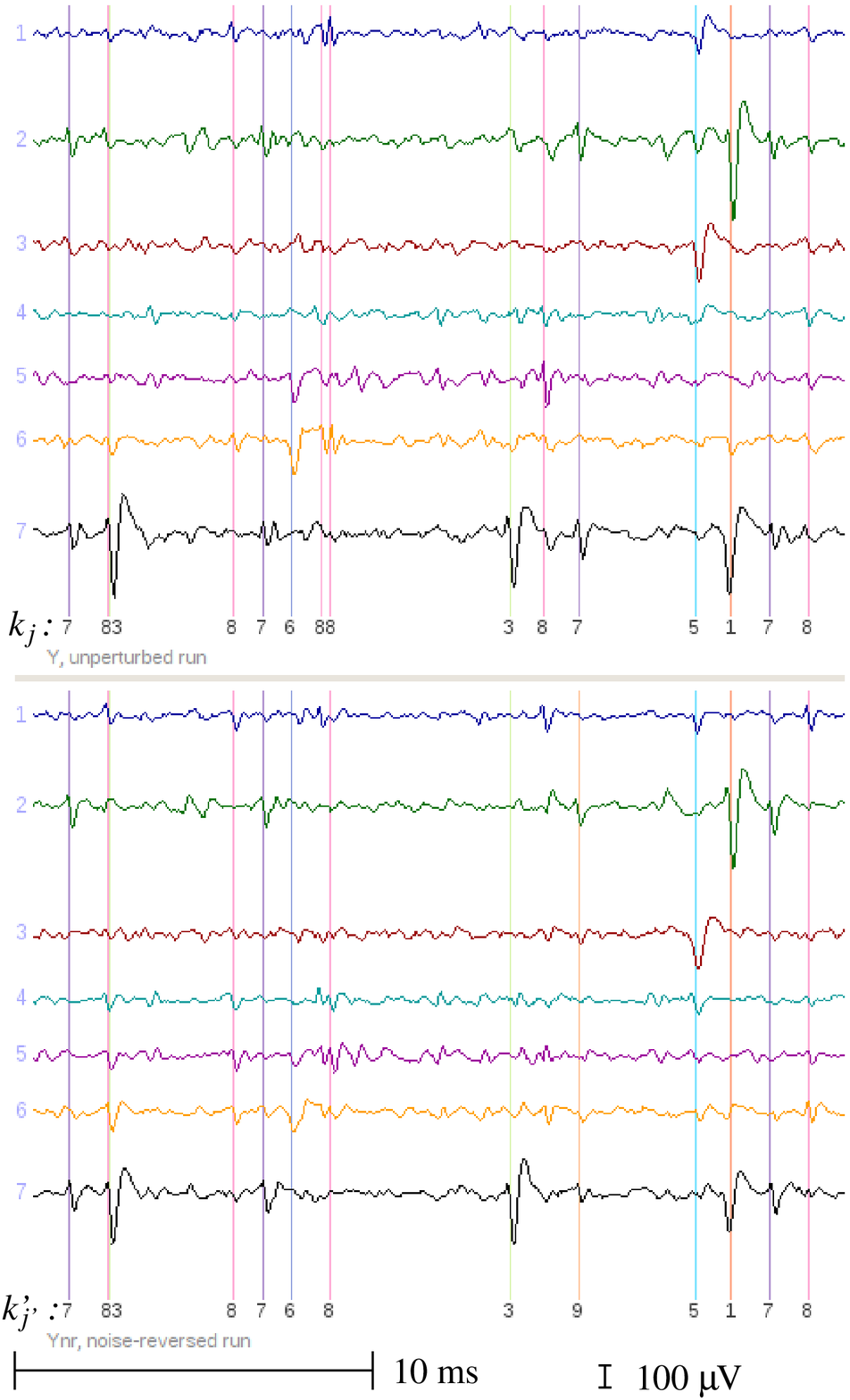}
\end{minipage}
\;
\begin{minipage}{2in}
  (e)\\
  \ig{width=2.1in}{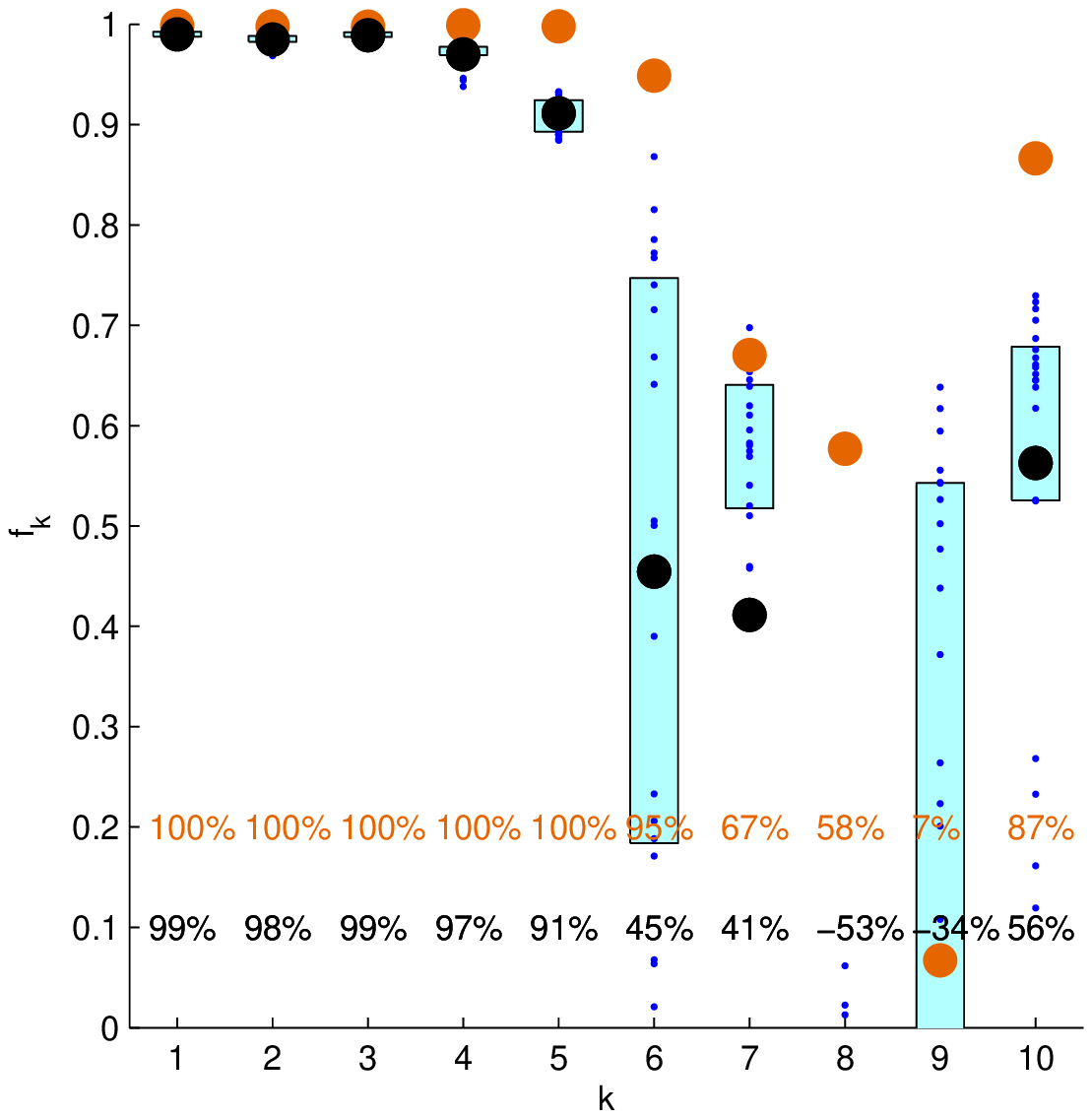}\\
  (f)\\
  \ig{width=2in}{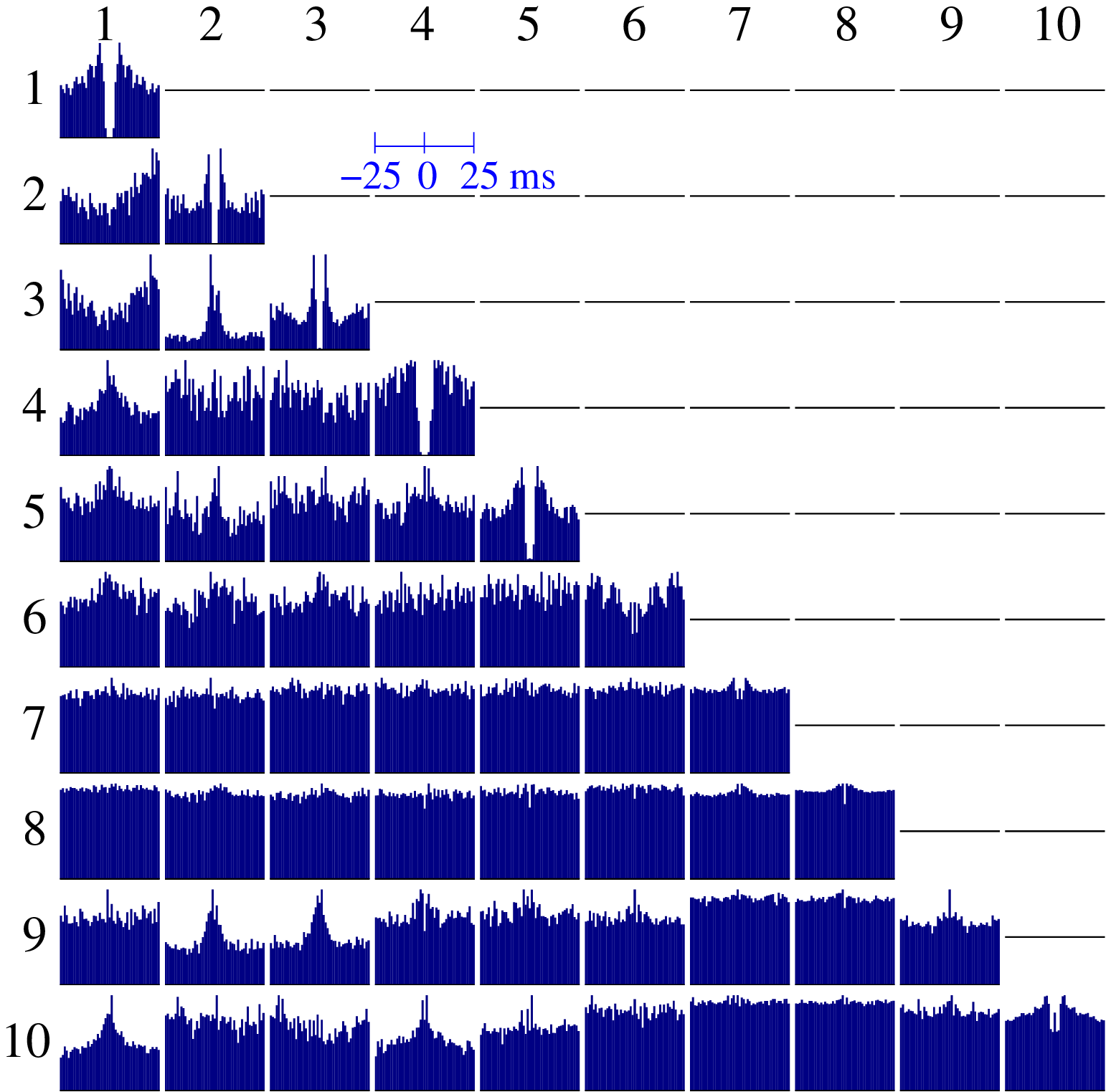}
\end{minipage}
\ca{Time-series stability metric results.
  (a) Toy example ($K=2$)
  of a matching $\mu$ between one spike sorter output
  $t_j$, $k_j$ and another $t'_{j'}$, $k'_{j'}$
  (blue shows label $k=1$ and red label $k=2$). The resulting
  extended confusion matrix and its ``best'' permutation $\hat{Q}$ are
  shown. The matching $\mu$ shown is the best matching $\hat\mu$
  even though the nearby pair is swapped in time.
  (b)-(f) show results for the spike sorter of
  \sref{s:tsalg} with $K=10$.
  (b) Waveform shapes $\{V^{(k)}\}_{k=1}^K$ and populations
  from the unperturbed run, and (c) from the noise-reversed run.
  (d) Input time series data $Y$, and noise-reversed time series $\tilde Y$,
  showing firing times and (best-permuted) labels from each run.
  (e) Stability metrics $f_k$ for noise-reversal (shown in orange), and for
  the spike addition metric (black).
  (f) Spike firing cross-correlations of spike sorter output $t_j$, $k_j$.
  Refractory dips (of at least $\pm 3$ ms) are complete for $k=1,\dots,5$ and partial for $k=6,10$, matching the metrics in (e).
}{f:tseries}
\efi
% add text after (b), (c), etc to describe...?

% aaaaaaaaaaaaaaaaaaaaaaaaaaaaaaaaaaaaaaaaaaaaaaaaaaaaaaaaaaaaaaaaaaaaaaaaaaaaa
\subsection{A simple time-series spike sorting algorithm}
\label{s:tsalg}

As an illustrative spike sorter we will use the following model-based
fitting algorithm which handles overlapping spikes
(the full description we leave for a future publication).
We assume that the time series $Y$ has been low-pass filtered.
Firstly, upsampled clips are extracted from $Y$ using
the triggering and upsampling procedure of App.~\ref{a:clips},
with a $-100$ $\mu$V threshold.
Secondly, these clips are fed into the clip-based spike sorter of
\sref{s:ssalg}, using the best of $r=100$ repetitions of k-means++,
with a preset number of clusters $K$.
Thirdly, a set of upsampled mean waveform shapes $\{W^{(k)}\}_{k=1}^K$
%(discarding the smallest-norm $K+1$th ``noise cluster'' waveform)
is found by applying \eqref{Wk} to the resulting clips and labels.
These waveforms are ordered by descending $l_2$-norm.
Finally, these waveforms are matched to the complete time series $Y$ via a
greedy fitting algorithm \cite{prentice,pillow}.
Specifically, the change in $l_2$-norm of the residual due to
subtraction of each (downsampled, fixed-amplitude)
waveform at each time-shift was
computed, and firing events $t_j$ were declared
at sufficiently negative local minima (wth respect to time shift)
of this function, that were also global minima over
all waveforms with time shifts up to $\pm 0.5$ ms from $t_j$.
The corresponding $k_j$ was the label of the waveform responsible
for this largest decrease in residual.
This greedy sweep through the time series was repeated
until no further change occurred.
%The algorithm parameters are thus $K$ and the prior firing rates.
Note that, as with many modern algorithms
\cite{takahashiICA,Franke,chaitu,Moore-Kochlacs2014,pillow}
the main fitting stage involves no concept of ``clips'' or event windows.

As in \sref{s:rerun}, a simple rerunning validation metric would be
possible for time series. However, we present the
following two metrics as much more realistic indicators of stability.

% nnnnnnnnnnnnnnnnnnnnnnnnnnnnnnnnnnnnnnnnnnnnnnnnnnnnnnnnnnnnnnnnnnnnnnnnnnn
\subsection{Noise-reversal}
\label{s:trev}

The noise-reversal idea of \sref{s:rev} generalizes to time
series, handling overlapping spikes naturally, as follows.
Firstly, the sorter runs on the unperturbed data to give
$S(Y) = \{t_j,k_j\}_{j=1}^{N_s}$.
Since the interface does not output waveforms, they are next estimated
by averaging%
\footnote{In the case of high firing rates, a least-squares solution of
  a large linear system may provide an improved estimation
  \cite{pillow,chaitu}.}
windows from $Y$ centered on the output firing times $t_j$ for the
appropriate label, eg
\be
V^{(k)}_{mt} \;= \; \frac{1}{n_k} \sum_{j: k_j=k} y_{m,t_j+t-T/2} ~,
\qquad %\mbox{for }
m = 1,\dots,M, \; t = 1,\dots,T ~.
\label{V}
\ee
The ``forward model'' generating a (noise-free) time series from
a new set of firing times $\mbf{s} := \{s_j\}_{j=1}^N$ and labels
$\mbf{l}:= \{l_j\}_{j=1}^N$ is then $F(\mbf{s},\mbf{l})$, with entries
\be
F_{mt} = F(\mbf{s},\mbf{l})_{mt} \;:=\; \sum_{j=1}^N V^{(l_j)}_{m,t-s_j+T/2}
~,
\qquad %\mbox{for }
m = 1,\dots,M, \; t = 1,\dots,\ttot ~.
\label{F}
\ee
(We in fact use
%more complicated formulae than
variants of \eqref{V}--\eqref{F}
which achieve sub-sample firing time accuracy by
upsampling the waveforms $V^{(k)}$
as in App.~\ref{a:clips}, and building downsampling into $F$.)
Using this, a perturbed time series (analogous to \eqref{rev})
is generated,
\be
\tilde Y  \;=\; 2F(\mbf{t},\kk) - Y
~,
\label{trev}
\ee
and sorted to give $S(\tilde Y) = \{t'_{j'},k'_{j'}\}_{j'=1}^{N_s'}$.
Finally, the best extended confusion matrix $\hat Q$
between the two sets of outputs is found as in \eqref{Qmu}--\eqref{Qextbest},
and the usual stability metric $f_k$ computed via \eqref{fk}.

%%%%%%%%%% Need to permute Wnr subfigure (c).
% Also, no splitting of k=1 now ???

\fref{f:tseries} illustrates this metric for spike sorting
filtered time series data $Y$
taken from an {\em ex vivo} retinal multi-electrode array recording
described in App.~\ref{a:tseries}, with $K=10$ neurons assumed.
This dataset has a large fraction of overlapping events:
spike sorting as in \sref{s:tsalg} results in around 10\% of firing
events falling within 0.1 ms of another event.
The upper half of \fref{f:tseries}(d) shows a short time window from $Y$,
while the lower half shows the resulting $\tilde Y$.
Note that some of the sorted times and labels have changed,
while between firing events the signal is {\em negated}.
The orange dots in \fref{f:tseries}(e) show the resulting $f_k^\tbox{rev}$
values: the five neurons with largest waveform norms are quite stable,
while $k=6$ and $10$ are marginally stable (recalling
from \sref{s:rev} that $90\%$ is hardly a stable noise-reversal metric),
and $k=7,8,9$ are unstable and---judging by their huge $n_k$---massively overfit.
This matches well the per-neuron
physiological validation based on clarity of refractory dips in
the firing time auto-correlations in \fref{f:tseries}(f).

It is worth examining why $k=9$ has almost zero stability. The cause is apparent
in panels (b)-(c), which show the waveforms $V^{(k)}$ estimated via \eqref{V}
from the unperturbed
and noise-reversed runs (the latter in ``best permuted'' ordering $\hat\pi$).
The two sets of waveforms are very similar apart from
the possibly distinct neuron $k=9$ in (b) which
has disappeared in (c) due to the splitting of $k=7$.
We find that such clustering instabilities are common, and vary
even with rerunning the sorter with a different random seed in k-means++.
Our metric quantifies these instabilities as induced by variations
consistent with any even-symmetric noise model.

% xxxxxxxxxxxxxxxxxxxxxxxxxxxxxxxxxxxxxxxxxxxxxxxxxxxxxxxxxxxxxxxxxxxxxxxxxxx
\subsection{Auxiliary spike addition}
\label{s:add}

The final stability metric we present
perturbs the time series by linearly adding
new spikes at known random firing times and then measuring their accuracy
upon sorting;
this gauges performance in the context of overlapping spikes---%
by creating and assessing new overlaps---while respecting the linearity of
the physical model for extracellular signals \cite{lfporigin}.

Let the first uperturbed sorter run produce times and labels
$S(Y) = \{t_j,k_j\}_{j=1}^{N_s}$, from which a forward model is
built, as above via \eqref{V}--\eqref{F}.
An estimate for the firing rate of the $k$th neuron is $n_k/\ttot$, where $n_k$
is
%its population
given by \eqref{nk}.
One generates an auxiliary set of events
$\{s_j,l_j\}_{j=1}^{N_a}$ as the union of $K$ independent
random Poissonian spike trains, with rates $\beta n_k/\ttot$,
$k=1,\dots, K$, where $\beta$ is a rate scaling parameter.
If $\beta$ is too small,
not much is learned from the perturbation;
% actually, any rerun-type instability causes large negative fk values
%
if too large, the total firing rates are made unrealistically large.
We fix $\beta=0.25$.
The perturbed time series is then
\be
\tilde Y \; = \; Y + F(\mbf{s},\mbf{l})
~,
\label{add}
\ee
which is sorted to give $S(\tilde Y) = \{t'_{j'},k'_{j'}\}_{j'=1}^{N_s'}$.
The best extended confusion matrix 
$\hat Q$ is now found (as in \eqref{Qmu}--\eqref{Qextbest})
between the total of $N_s+N_a$ events
$\mbf{t} \cup \mbf{s}, \,\kk \cup \mbf{l}$,
and the second run outputs $\mbf{t}',\kk'$.
To assess the {\em change} in $\hat Q$ induced by adding spikes,
we compute the matrix
$$
\hat Q^\tbox{add} \;:=\; \hat Q - \diag\{n_1,\dots,n_K,0\}
$$
where $\diag$ denotes the diagonal matrix with diagonal elements as listed.
Finally, $f_k^\tbox{add}$ is computed as in \eqref{fk} using $\hat Q^\tbox{add}$.
One may interpret this diagonal subtraction as follows:
if all the new spikes were correctly sorted, and all the original
spikes sorted as in the first run,
$\hat Q^\tbox{add}$ would be diagonal with
entries given by the added populations for each label,
so $f_k^\tbox{add} = 1$ for all $k$.
Thus low $f_k^\tbox{add}$ values indicate either lack of sorting accuracy
for the added spikes,
or induced instability (eg false positives or negatives)
in re-sorting the original spikes.
Both of these are important warning signs that a
putative neuron is less believable.

\begin{rmk}
Marre et al.\ \cite{salamander} and
Rossant et al.\ \cite{rossant} use similar but distinct validation metrics:
they add spikes with known firing times
from either a spatially translated neuron, or a
``donor'' neuron from a different recording, then
assess accuracy for this neuron.
While useful as overall measures of reliability,
these methods require dataset-specific adaptations, and
it is not clear that they validate any of the particular neurons
{\em in the original data}.
In contrast, the metric we propose here % in this section
%provides a %per-neuron
validates each neuron as sorted in the dataset of interest.
\end{rmk}

\fref{f:tseries}(e) shows 20 independent samples
of $f_k^\tbox{add}$ (ie, 20 realizations of the Poisson spike trains),
with the black dots showing the means.
Only the first five neurons have stabilities above 60\%.
The overall pattern is similar to noise-reversal,
but with a lower overall calibration of the metric,
and it matches well the independent validation via cross-correlations
in \fref{f:tseries}(f).
Note that if adding a certain type of spike induces changes in the sorting
of the (more numerous) existing spikes, then huge drops in stability,
including highly negative $f_k^\tbox{add}$,
are possible, as visible in \fref{f:tseries}(e), $k=8,9$.
Indeed, from its waveform and huge population in \fref{f:tseries}(b),
$k=8$ would likely
be classified as the ``noise cluster'' (see eg \cite{Wild2012})
%
%*** give waveclus /osort ref?
%
by an experienced human.
% positive auto-corr spike in k=8 shows false positives?

%In summary, the spike addition metric provides more realistic
%stabilities than noise-reversal.

%We believe that this is important in assessing the inducing of
%false positive events.

%Note the smallest l2 norm Ws are not always the least stable. ($k=10$).

\begin{table} % ttttttttttttttttttttttttttttttttttttttttttttttttttttttttttttttt
{\hspace{-2ex} \scriptsize%
  \begin{tabular}{l|l|l|l|l|}
Interface& Metric & Section & Advantages & Disadvantages
\\
\hline
clip based& rerun & \sref{s:rerun} &
simplicity & useless if sorter is deterministic
\\
& 3-way CV & \sref{s:cv} &
simplicity &
classifier needed;
%run size only $N/3$;
not useful for large $N$
\\
& self-blurring & \sref{s:blur} &
adjustable parameter $\gamma$
& raises noise level (can underestimate accuracy)
\\
& noise reversal & \sref{s:rev} &
%also works for $t$-series.
only needs 2 runs; no extra noise &
stability range close to 1
\\
\hline
time series & noise reversal & \sref{s:trev} &
%also works for clips.
only needs 2 runs; no extra noise&
stability range close to 1;
reverses noise polarity
%&&&&unrealistic if noise not sign-invariant
\\
& spike addition & \sref{s:add} &
accuracy in context; no extra noise & firing rate slightly higher;
no built-in shape variation\\
\hline
  \end{tabular}
  }%
\ca{Summary of the proposed
validation metrics for clip-based and time series spike sorting
algorithms.}{t:sum}
\end{table}

% DDDDDDDDDDDDDDDDDDDDDDDDDDDDDDDDDDDDDDDDDDDDDDDDDDDDDDDDDDDDDDDDDDDDDDDDDDDD
\section{Discussion and conclusions}
\label{s:conc}

We have proposed several schemes for the validation of automatic
spike sorting algorithms, each of which outputs a per-neuron stability
metric in the context of the dataset of interest,
without ground truth information or physiological criteria,
and with few assumptions about the noise model.
Our key contribution is a common framework (\fref{f:inter})
where this validation occurs
only through a simple standardized interface to the sorter, which is
treated as a {\em black box} that can be invoked multiple times.
We see this as essential to automated
validation and benchmarking of a growing variety of
spike sorting algorithms and codes. % which have different internal workings.
% on an equal footing.
By contrast, most existing validation metrics rely, in part,
on accessing internal parameters peculiar only to certain algorithms.
We also envisage such metrics as the first filter
in a laboratory pipeline in which---%
due to the increasing scale of multi-electrode recordings---%
it will be impossible for a human operator to make decisions about most neurons.
% as appreciated in the clustering literature,\cite{clusterwise,binyu2013})
While stability cannot guarantee accuracy
or single-unit activity (eg see \sref{s:perttest}),
instability is a useful warning of likely inaccuracy or multi-unit activity.
Only neurons that pass the stability test should
then be further validated based on eg shape \cite{tankus}, refractory gaps
\cite{Hill2011} and/or receptive fields \cite{litke,salamander,pillow}.

Table~\ref{t:sum} summarizes the six schemes: the first four
use a simple sorting interface that classifies ``clips'' (spiking events),
while the remaining two use a more general interface that
accesses the full time series and allows handling of overlapping
spikes.
All schemes resample or perturb the data in a realistic way, then
measure how close the resulting neuron-to-neuron {\em best confusion matrix} $\hat Q$
is to being diagonal; its off-diagonal structure also
indicates which neurons are coupled by instabilities.
Although our demonstrations used standard sorting algorithms with a specified
number of neurons $K$, the metrics also apply to variable $K$ with minor extra
bookkeeping.

The first two schemes (rerunning and cross-validation)
are conceptually simple but are shown to have severe limitations.
The next two (self-blurring and noise reversal for clip sorting) seem useful even
in large-scale settings.
However, we
expect that the last two (noise reversal and spike addition for time series)
%time series interface metrics
will be the most useful.
Each metric carries its own {\em calibration}:
for instance, with the sorter fixed, for noise reversal $f_k$ is much closer to 1 than for spike addition.
A future task is to calibrate the metrics against accuracy with
both synthetic and ground-truth data.

Although general, our schemes make certain assumptions.
Noise reversal assumes sign-symmetry in the noise distribution
(which may not be satisfied for ``noise'' due to distant spikes),
whereas spike addition as presented does not account for realistic
{\em firing variation} in the
added spikes; the latter could be achieved in a model-free way by
adding random spike instances (as in self-blurring) instead of mean waveforms.
% but this would increase the noise level. Cannot separate measurement or
% distant firing noise from firing-varation of spike.

%Besides this,
We believe that several other extensions are worth exploring.
1) The last two validation schemes essentially involve reverse-engineering
a forward model \eqref{F} from a single spike sorter run.
This model could be improved in various ways, by making use of
firing amplitude information (if provided by the sorter),
%The interface could be ``overloaded'' to make use of firing amplitude (or
or, in nonstationary settings (waveform drift),
constructing waveform shapes from only a local moving window.
2) The random known times in spike addition could be chosen to respect
the refractory periods of the existing spikes, avoiding
rare but unrealistic self-overlaps.
%that can currently happen.
Added spikes could also be chosen in a bursting
sequence that tests the sorter's accuracy in this difficult setting.
3) A variant of self-blurring that estimates
$\partial f_k^\tbox{blur}/\partial \gamma |_{\gamma=0}$ could avoid the
problem of noise level increase (cluster inflation) at larger $\gamma$.
%avoids making clusters larger, but requires many more sorting runs.
%4) For nonstationary {\em in vivo} settings, waveform shapes used in the
%forward model \eqref{F} could be taken only from a local moving window.
% no model of waveform drift needed

Code which implements the validation metrics and sorting algorithms used
is freely available
(in MATLAB with a MEX interface to C) at the following URL:

{\tt https://github.com/ahbarnett/validspike}
  
% Fee1996 uses clips but also their time info to use xcorr to merge clusters
% Allow t_j input along with each clip?

%We hope that the presented validation metrics 

%Self-blurring is a measure of point density lying near decision
%boundaries, but without reference to feature space.
%The concept of ``near'' is determined here by the variation in the
%dataset itself; no external noise model is needed
%(unlike, eg, $L_\tbox{ratio}$ of \cite{Schmitzer-Torbert2005}).

%could extract more info from Qind re types of induced false positives

%could add variation or add random events from $Y$ as in self-blurring,
%but would add more noise.

%We did not test clip-based sorting algorithms that choose their own $K$,
%but each clip-based validation scheme we present is compatible with that.
%(simple bookkeepping)

%noise reversal with scaling?

%For simplicity, fixed firing amplitudes are assumed in noise rev
%and spike addition.

%For addition, could choose a random other spike of the desired type,
%which would provide natural variation without an explicit
%model or assumptions about the variation. But would also add noise.

%treat firing amplitude \& width as extra outputs and use for fwd model run?

%One obvious extension is to validation of one spike sorting algorithm against
%another, or against the same algorithm with different parameters.
%We expect this to be useful. 
%We leave this for future work.

\section*{Acknowledgments}

We are very grateful for the Chichilnisky and Buzs\'aki labs for supplying
us with test datasets.
We have benefited from many helpful
discussions, including those with
EJ Chichilnisky, Gy\"orgy Buzs\'aki,
Adrien Peyrache, Brendon Watson, Bin Yu, Eero Simoncelli,
Mitya Chklovskii, and Eftychios Pnevmatikakis.

\appendix % AAAAAAAAAAAAAAAAAAAAAAAAAAAAAAAAAAAAAAAAAAAAAAAAAAAAAAAAAAAAAAAAA
\section{Provenance of test datasets}

\subsection{Clips dataset}
\label{a:clips}

We started with an electrical recording,
supplied to us by the Buzs\'aki Lab at NYU Medical Center,
taken from the motor cortex of a
freely behaving rat at
20 kHz sampling rate with a NeuroNexus ``Buzsaki64sp'' probe
\cite{brendon}.
%\cite{
We took $3\times 10^6$ consecute time
samples from a single shank with $M=10$ channels.
% gp 5
The following procedure to extract clips combines standard techniques
in the literature.

We first high-pass filtered at 300 Hz, using the FFT and a frequency filter
function $a(f) = (1+\tanh[(f-300)/100])/2$, where $f\ge 0$ is the frequency
in Hz.
This gives a $M$-by-$3\times 10^6$ matrix $Y$,
which was then spatially prewhitened by replacing by $QY$, where
$Q$ is $R^{-1}$ but with each row normalized to unit $l_2$ norm.
$R^T R = C$ is the Cholesky factorization of $C$, an estimate of
the $M\times M$ channel-wise cross-correlation matrix.
$C$ itself was estimated using $10^4$ randomly-chosen ``noise clips''
(time-series segments of duration 4 ms
for which no channel exceeded a threshold of 158 $\mu$V).
This prewhitening significantly improved the noise level,
and helped remove common-mode events.
Events were triggered by the minimum across channels passing below $-120$ $\mu$V
(around 4 times the estimated noise standard deviation);
events where this trigger lasted longer than 1 ms, or where another trigger
occurred within 2 ms either side of the triggering event, were discarded
as unlikely to be due to a single spiking event.
This gave 6901 clips of length 60 samples each.

Clips were then upsampled by a factor of 3, using interpolation from the
sampling grid by the Hann-windowed sinc kernel \cite{blanche}
of width $\tau = 5$ samples,
$$
f(t) = \left\{\begin{array}{ll}
  1, & t = 0, \\
  0, & |t|>\tau, \\
  \frac{\sin(\pi t)}{\pi t} \, \cos^2 \bigl( \frac{\pi t}{2 \tau}  \bigr),&
  \mbox{otherwise},
  \end{array}\right.
$$
where $t$ is given in the original sample units.
The kernel width reduced the clip duration from 3 ms to 2.45 ms.
Finally, the upsampled clips were aligned by time translation by an integer
number of upsampled grid points until the minimum across channels
lies at the central grid point.
When translated, values either side of the data were padded with zeros
(on average less than one padded zero per clip per channel was needed).
The upsampling allows alignment to be more accurate
and improves cluster quality \cite{blanche}.
%and prevents false splitting of clusters.
The result was a $M$-by-$T$-by-$N$ array of clips $X$ with $M=10$, $T=147$,
and $N=6901$.
%The code for this is available at {\tt https://github.com/ahbarnett/spikesort}
%etc

\subsection{Larger clips dataset}
\label{a:large}

In Sections~\ref{s:cv} and \ref{s:pert}
datasets comprising a larger number of clips $N$ were needed.
Since we did not have stationary time series containing such large numbers
of clips, we generated them from the above clips as follows.
Let $F$ be the factor to grow the number of clips.
Each clip was duplicated $F$ times, then to each channel of each new clip was
added time-correlated Gaussian noise with variance $\eta^2$, where
$\eta = 30$ $\mu$V is an estimate of the noise standard deviation from
pre-whitened ``noise clips'', as above.
The autocorrelation function in time was taken as $C(t) = e^{-t/\tau}$,
with $\tau = 0.5 $ ms, giving a rough approximation to the observed
noise autocorrelation.
As is standard, the time-correlated noise samples were generated by
applying the Cholesky factor of the time-sample autocorrelation matrix
to an iid Gaussian signal of the desired length.

\subsection{Time-series dataset}
\label{a:tseries}

Our time-series data $Y$
comes from $M=7$ adjacent electrodes (spatially forming a
hexagon around a central electrode) within a 512-electrode
array \cite{litke} recording at a 20 kHz sample rate from
the spontaneous activation of an {\em ex vivo} monkey retina. % taken in 2005.
The recording length was 2 minutes ($T_\tbox{tot} = 2.4 \times 10^6$ samples),
and mean neuron spiking rates are of order 20 Hz.
This was supplied to us by the Chichilnisky Lab at Stanford.
The raw data was then high-pass filtered at 300 Hz as in
App.~\ref{a:clips}. Since there was little noise correlation between channels,
no spatial prewhitening was done.

\section{Analytic stability for a Gaussian erroneously split in two}
\label{a:gauss}

When the $N$ clips are drawn from some underlying
probability distribution function (pdf) $p$ in signal or feature space,
we have analytic results as $N\to\infty$ for the stability of a single cluster
when split into two symmetrically by a decision hyperplane
(eg by k-means with $K=2$).
Consider as in \fref{f:pert}(d), the 1D case
$z\in\RR$, with $p(z)$
% even-
symmetric about the splitting
point $z=0$ (without loss of generality).
The centroid of the positive half of the pdf is
$c = 2 \int_0^\infty z p(z) dz$.
The stability (of either of the two labels)
is the mass that does not change label,
which, specializing to $\gamma=1$, is the chance that the sum of
two independent positive samples from $p$ exceeds $c$,
\be
f_1^\tbox{blur} = 1 - 4 \int_0^c \int_0^{c-y} p(x)p(y) dx \,dy~.
\label{f1blur}
\ee
Choosing a Gaussian $p(z) = (1/\sqrt{\pi}) e^{-z^2}$ for which
$c = 1/\sqrt{\pi}$,
the second term in \eqref{f1blur}
is the mass in a diamond region which, by the rotational
invariance of $p(x)p(y)$, may be rotated by $\pi/4$ to
become the square $|x|,|y|<c/\sqrt{2}$.
Thus
$$
f_1^\tbox{blur} = 1 - \biggl[ 2 \int_0^{c/\sqrt{2}} p(x) dx \biggr]^2
= 1 - [ \erf(1/\sqrt{2\pi})]^2~ = 0.817\dots ,
$$
where the usual error function is defined in \cite[(7.2.1)]{dlmf}.

%Self-blurring replaces the positive half of the pdf $p$ with the
%convolution of this
%positive half with a $\gamma$-scaled and $c$-centered version of itself,
%ie
%\be
%\tilde p(z) = 4 \int_0^{\gamma c + z} p(x) p\bigl(\frac{z-x}{\gamma} + c\bigr) %dz
%\ee

For noise reversal, stability is simply the probability that a single
sample has $|z|<2c$,
$$
f_1^\tbox{rev} = \erf(2c) = \erf(2/\sqrt{\pi}) \approx 0.889\dots
$$
Both of these cases apply to a multivariate Gaussian split on a
symmetry hyperplane, by taking $z$ as the
coordinate normal to the hyperplane.

% BBBBBBBBBBBBBBBBBBBBBBBBBBBBBBBBBBBBBBBBBBBBBBBBBBBBBBBBBBBBBBBBBBBBBBBBBBBB
\bibliographystyle{abbrv}
\bibliography{spikesorting}
\end{document}